\documentclass{article}

\usepackage{microtype}
\usepackage{graphicx}
\usepackage{subcaption}
\usepackage{booktabs} 
\usepackage{multirow}
\usepackage{fancyvrb}
\usepackage{hyperref}
\usepackage{amssymb}
\usepackage{adjustbox}
\usepackage{booktabs}
\usepackage{amsfonts}
\usepackage{amssymb}
\usepackage{hyperref}
\usepackage{xcolor} 
\usepackage{colortbl}
\usepackage{graphicx}
\usepackage{array}
\usepackage{multirow}
\usepackage{wrapfig}
\usepackage{float}
\usepackage{caption}
\usepackage{rotating}
\usepackage{enumitem}
\usepackage{makecell}
\usepackage{subcaption}
\usepackage{marvosym}
\usepackage{verbatim}
\usepackage{listings}
\usepackage{titletoc}
\usepackage[most,listings,breakable]{tcolorbox}
\usepackage{pifont}
\newcommand{\cmark}{\checkmark}
\newcommand{\xmark}{\ding{55}}
\usepackage{microtype}      
\usepackage{xcolor}         
\usepackage{times}
\usepackage{latexsym}
\usepackage{wrapfig}
\usepackage{graphicx}
\usepackage{algorithm}
\usepackage{algpseudocode}
\newcommand{\TopM}{\mathrm{TopM}}
\newcommand{\TopK}{\mathrm{TopK}}
\usepackage{float}
\usepackage{booktabs} 
\usepackage{longtable} 
\usepackage{arydshln}
\newcommand{\rcomment}[1]{\hfill$\triangleright$~{\footnotesize #1}}
\usepackage{pgf}
\usepackage{pifont}
\usepackage{url}
\usepackage{longtable}
\usepackage[utf8]{inputenc} 
\usepackage[T1]{fontenc}    
\usepackage[preprint]{icml2026}
\usepackage{booktabs}

\usepackage{multirow}
\usepackage{amsmath}
\usepackage{amssymb}
\usepackage{mathtools}
\usepackage{amsthm}

\usepackage[capitalize,noabbrev]{cleveref}

\theoremstyle{plain}

\theoremstyle{definition}

\theoremstyle{remark}

\usepackage[textsize=tiny]{todonotes}
\newcommand{\gain}[1]{%
  \begingroup
  \pgfmathparse{#1<5 ? 1 : (#1<10 ? 2 : (#1<20 ? 3 : 4))}%
  \edef\bucket{\pgfmathresult}%
  \ifnum\bucket=1 \textsuperscript{\textcolor{green!40!black}{\scriptsize +#1}}%
  \else\ifnum\bucket=2 \textsuperscript{\textcolor{green!70!black}{\scriptsize +#1}}%
  \else\ifnum\bucket=3 \textsuperscript{\textcolor{green!90!black}{\scriptsize +#1}}%
  \else                 \textsuperscript{\textcolor{teal!80!black}{\scriptsize +#1}}%
  \fi\fi\fi
  \endgroup
}

\begin{document}

\twocolumn[
  \icmltitle{TARSE: Test-Time Adaptation via Retrieval of Skills and Experience for Reasoning Agents}



  \icmlsetsymbol{equal}{*}

  \begin{icmlauthorlist}
    \icmlauthor{Junda Wang}{yyy}
    \icmlauthor{Zonghai Yao}{yyy}
    \icmlauthor{Hansi Zeng}{yyy}
    \icmlauthor{Zhichao Yang}{yyy}
    \icmlauthor{Hamed Zamani}{yyy}
    \icmlauthor{Hong Yu}{xxx}
  \end{icmlauthorlist}

  \icmlaffiliation{yyy}{University of Massachusetts Amherst, MA, USA}
  \icmlaffiliation{xxx}{University of Massachusetts Lowell, MA, USA}
    \icmlcorrespondingauthor{Junda Wang}{jundawang@umass.edu}
  \icmlkeywords{Machine Learning, ICML}

  \vskip 0.3in
]



\printAffiliationsAndNotice{}  

\begin{abstract}
Complex clinical decision making often fails not because a model lacks facts, but because it cannot reliably select and apply the right procedural knowledge and the right prior example at the right reasoning step. 
We frame clinical question answering as an agent problem with two explicit, retrievable resources: skills, reusable clinical procedures such as guidelines, protocols, and pharmacologic mechanisms; and experience, verified reasoning trajectories from previously solved cases (e.g., chain-of-thought solutions and their step-level decompositions). 
At test time, the agent retrieves both relevant skills and experiences from curated libraries and performs lightweight test-time adaptation to align the language model’s intermediate reasoning with clinically valid logic. 
Concretely, we build 
(i) a skills library from guideline-style documents organized as executable decision rules, 
(ii) an experience library of exemplar clinical reasoning chains indexed by step-level transitions, and 
(iii) a step-aware retriever that selects the most useful skill and experience items for the current case. 
We then adapt the model on the retrieved items to reduce instance–step misalignment and to prevent reasoning from drifting toward unsupported shortcuts. 
Experiments on medical question-answering benchmarks show consistent gains over strong medical RAG baselines and prompting-only reasoning methods. 
Our results suggest that explicitly separating and retrieving clinical skills and experience, and then aligning the model at test time, is a practical approach to more reliable medical agents.~\footnote{The code will be released after accept.}
\end{abstract}

\begin{figure*}[!ht]
    \centering
    \includegraphics[width=0.96\linewidth]{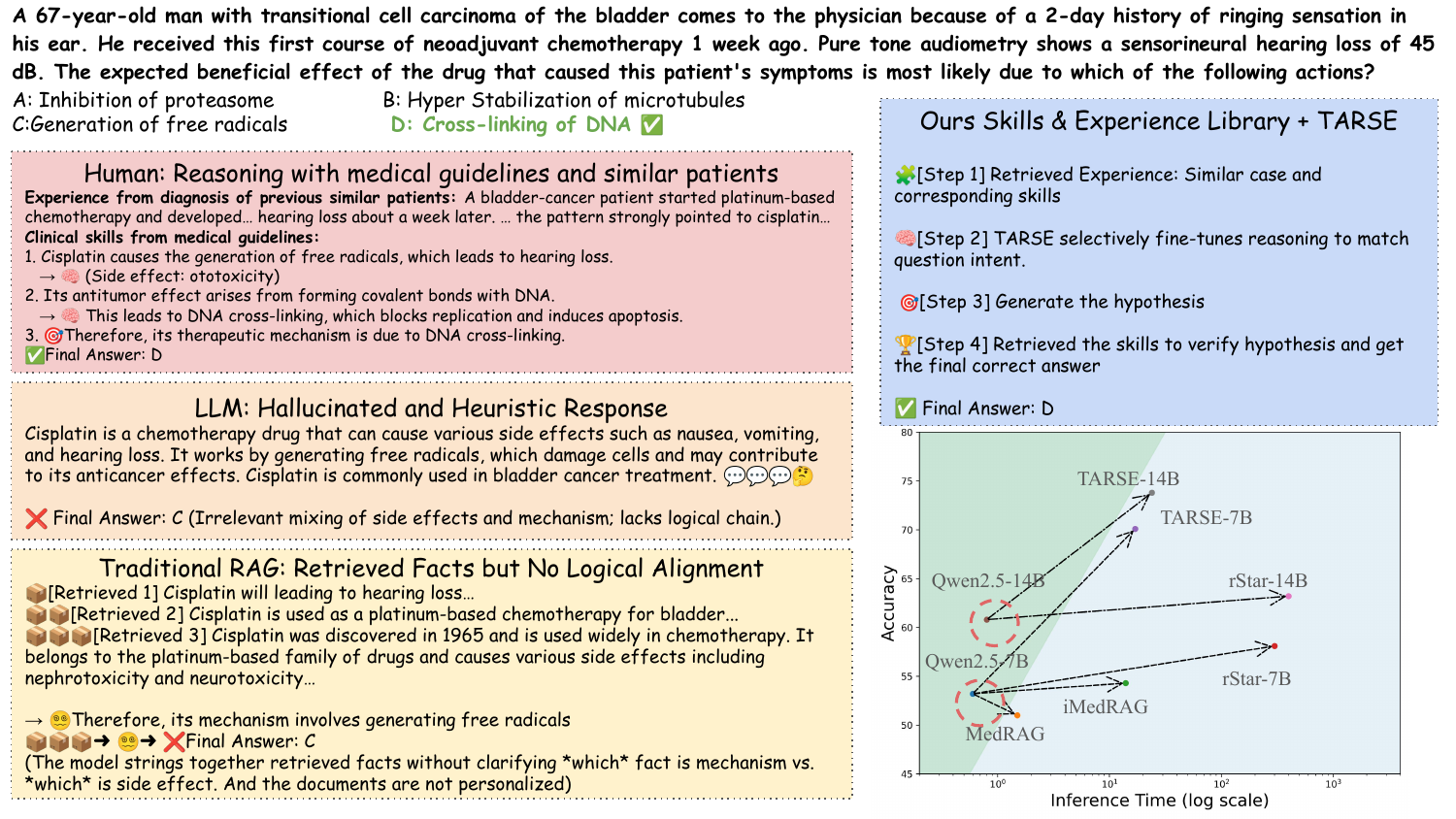}
    \caption{
       \textbf{Motivation and overview of Skills \& Experience Library + TARSE}.
       \textbf{(Left)} A clinical QA example where the key difficulty is \emph{goal-conditional disambiguation}: cisplatin’s \emph{ototoxicity} (free radicals $\rightarrow$ hearing loss) is a side-effect chain, while its \emph{antitumor mechanism} is the therapeutic chain (covalent DNA binding $\rightarrow$ DNA cross-linking $\rightarrow$ apoptosis). Humans solve this by combining experience (recalling a similar bladder-cancer + platinum-chemo case that links the symptom to cisplatin) with skills (reusable guideline/mechanism logical chains) to apply a “mechanism vs. side-effect” gate and select the correct branch (DNA cross-linking, option D). 
       In contrast, a vanilla LLM tends to follow heuristic associations and conflates toxicity with mechanism, and traditional chunk RAG retrieves true but unstructured snippets that are not step-aligned, leading to the same confusion.
       \textbf{(Right)} Our method operationalizes this workflow: (1) retrieve similar experience traces and relevant skills; (2) perform lightweight test-time adaptation to align reasoning with the question intent; (3) generate a provisional hypothesis/chain; and (4) verify/repair it using retrieved skills to produce the final answer. 
       The bottom-right plot summarizes the accuracy–latency trade-off, showing that TARSE achieves larger gains than CoT-only and conventional RAG baselines at comparable inference time.}
    \label{fig:motivation}
    \vspace{-6mm}
\end{figure*}

\section{Introduction}
Clinical decision questions often look simple, but they hide a hard branching rule~\citep{jiang2025hulu,sellergren2025medgemma,xu2025lingshu}. For example, a question may ask what to do next for a heart-attack-like presentation at a small hospital, where a fast “clot-busting” option exists, but the patient also has a recent brain-bleed history~\citep{li2023chatdoctor, wang2025huatuo, wang2024notechat}. 
A general LLM with chunk-based RAG can retrieve several true snippets (e.g., “act fast” and “bleeding risk”)~\citep{jin2025search,zhao2025medrag,xiong2024improving,zakka2024almanac,lozano2023clinfo,wang2024jmlr}, but it can still fail because it does not treat the hard rule as a gate at the right step. In contrast, a medical agent that can retrieve (i) a procedural skill that explicitly encodes the branch rule (including a “do-not-use” condition), and (ii) a similar verified experience trace that checks this condition before choosing the branch, can follow the correct decision path (as shown in Figure~\ref{fig:motivation}).

This example illustrates two resources that clinicians repeatedly rely on. First, they use reusable procedural knowledge (e.g., guidelines, protocols, decision rules, and pharmacologic mechanisms) that can be applied across many cases. Second, they draw on concrete prior experience (e.g., how similar cases were solved step by step, including common pitfalls and how constraints were handled). We formalize these resources for medical agents. In this work, we define clinical skills as structured procedural knowledge that can be reused across patients. We define clinical experience as verified reasoning trajectories from previously solved cases, represented as exemplar chains of thought and their step-level decompositions that expose which inference or check was required at each step. Treating both as explicit retrieval targets enables an agent to call on them at test time to guide multi-step inference.

Despite rapid progress in medical LLMs and retrieval-augmented generation~\citep{wu2024medical,wang2024jmlr,jiang2025tc}, current systems rarely support this skill–experience view~\citep{park2025instruct}. First, most medical RAG pipelines retrieve unstructured document chunks, which blur the distinction between procedural rules and case-specific reasoning traces, making it unclear what the model should retrieve to make a decision~\citep{xia2025medrek,liang2025rgar,myers2025evaluating}. Second, even when relevant information is retrieved, it is often misaligned with the critical reasoning step: multi-hop clinical questions frequently hinge on a single missing check (e.g., contraindications, time windows, or exclusion criteria), and global retrieval does not reliably inject the right rule at the right point. Third, dominant medical RAG approaches (e.g., MedRAG-style~\cite{xiong2024benchmarking} or iterative chunk retrieval~\cite{xiong2024improving}) are not designed to retrieve skills and experience as structured objects~\citep{ou2025experience}. Guidelines are executable procedures, and reasoning chains are trajectories; fragmenting them into generic chunks can destroy their decision structure and encourage the model to “average” across partially relevant snippets rather than following a coherent clinical pathway. These limitations motivate a test-time agent that retrieves structured skills and experience and aligns the model’s reasoning with them step by step.

We propose a retrieval-augmented medical agent that explicitly separates and operationalizes clinical skills and experience. We construct a skills library from guideline-like documents organized into reusable decision procedures, and an experience library of exemplar clinical reasoning chains indexed by step-level transitions. At inference time, the agent performs step-aware retrieval over both libraries, selecting the most relevant skill items and experience items. We then perform lightweight test-time adaptation on the retrieved items to align the language model’s intermediate reasoning with clinically valid logic, reducing instance–step misalignment and preventing reasoning from drifting toward unsupported shortcuts. We evaluate our approach on medical question-answering benchmarks and compare against strong medical RAG baselines and prompting-only reasoning methods. Across tasks, we observe consistent improvements, with larger gains on multi-hop and constraint-heavy questions that require explicit checks and branching decisions.

\noindent\textbf{Contributions:}
i) we introduce a skill–experience framing for clinical reasoning and argue that chunk-based retrieval is structurally mismatched to these resources;
ii) we build reusable libraries of clinical skills and exemplar experience with step-level indexing for retrieval;
iii) we present a test-time retrieval-and-adapt framework that improves medical reasoning performance and yields more interpretable, clinically grounded decision paths.

\section{Related Work}

\noindent\textbf{Skills and Experience}
In the agent literature, \emph{skills} are reusable procedures that an agent can call across tasks or states, and many works study how to discover and reuse them through hierarchical control, skill induction, and libraries of callable routines \citep{park2025instruct,wang2025inducing}. Skills matter because they turn a long task into shorter, stable subroutines, so an agent can generalize by selecting and composing what it already knows. A related line of work stores skills as retrievable artifacts at inference time, such as tool-use templates, program sketches, natural-language procedures, or checklists \citep{jin2025search,xiong2024improving,xia2026skillrl}. This matches expert work in domains with written rules. In medicine, guidelines and protocols serve this role, since they encode decision steps and exclusion rules meant to be reused across patients. Prior systems retrieve curated guidance \citep{zakka2024almanac}, but most work is prompt-centric and treats these documents as general evidence rather than as structured skills that govern step-by-step decisions.

Our method follows the skill-library view and makes it concrete for clinical reasoning. We represent guideline-like procedures as structured skill items (e.g., required checks, branch conditions, exclusion criteria) and retrieve them as objects that specify the next step, rather than as loose text. We also integrate skills with \emph{experience} in a single test-time loop. In reinforcement learning, replay and off-policy methods reuse trajectories to stabilize learning and transfer \citep{zhang2025agent}, and recent work also leverages self-generated trajectories and their outcomes as supervision \citep{wang2025steca,hu2025agentgen}. For LLMs, chain-of-thought and memory-based methods reuse reasoning traces to guide new problems \citep{wei2022chain,yan2025memory}, but traces are usually retrieved as extra context without an explicit procedure, so models can retrieve correct facts yet execute the wrong step.

We address this by curating \emph{experience} as verified step-by-step traces and indexing them by step-level transitions, so retrieval targets the missing step rather than only overall similarity. At test time, we retrieve both (i) the relevant skill items from guidelines and (ii) matched trace segments, then align the model step by step with the retrieved skill items, using traces to show how checks were executed in similar cases. This targets a common failure in chunk-based RAG: retrieving true statements but omitting a required check or applying a hard rule at the wrong point.

\noindent\textbf{Reasoning with LLMs and Retrieval}
\label{subsec:rw_reasoning_rag}
LLMs still struggle with explicit multi-step inference and check-and-branch reasoning~\citep{liu2023evaluating,wu2024reasoning,dziri2023faith}. Prompting methods (e.g., CoT, self-consistency) can improve fluency~\citep{wei2022chain,wang2022self}, but they do not reliably enforce correct step order or verifiable decisions in expert domains. RAG reduces unsupported claims by injecting external evidence~\citep{lewis2020retrieval,guu2020retrieval}, yet retrieved text is often unstructured or mismatched to the specific step that needs support~\citep{ye2023natural,xiong2024benchmarking}. Even medical RAG variants (e.g., MedRAG, iMedRAG) largely retrieve generic document chunks, which is insufficient when a single missed check flips the decision
Our work reframes retrieval targets for clinical reasoning: instead of retrieving only chunks, the agent retrieves explicit procedural skills (reusable rules and checklists) and exemplar experience traces (verified, step-by-step solutions). This makes retrieval actionable for decision-making and reduces step mismatch by providing the appropriate rule or example at the step where it is needed.

\noindent\textbf{Test-Time Training and Adaptation for Reasoning}
\label{subsec:rw_ttt}
Test-time training/adaptation updates a model \emph{during inference} using test-time signals; it is well studied in vision~\citep{sun2020test,behrouz2024titans} but remains limited for NLP and LLM reasoning~\citep{leera}. Prior objectives are often unsupervised and weakly aligned with the target reasoning process, so errors caused by \emph{instance--step misalignment} persist: the model retrieves relevant evidence at a question level but still applies the wrong rule at the wrong step.
We address this by coupling retrieval with lightweight test-time adaptation on the agent’s retrieved \emph{skills} and \emph{experience} items. The retrieved artifacts provide structured, instance-specific supervision that aligns the model’s step-by-step reasoning to the correct decision logic without global retraining.

\section{Problem Formulation}

\paragraph{Hidden-chain view of reasoning.}
Let $q\in\mathcal{Q}$ be a question and $M_\theta$ a (small) language model with parameters $\theta$.
We posit that answering proceeds through an \emph{unobserved} sequence of micro–inference states
$\mathbf{z}=\langle z_1,\ldots,z_T\rangle\in\mathcal{Z}^T$ (the \emph{latent logical chain}),
followed by emission of an answer $y\in\mathcal{Y}$:
\[
y \;=\; f_\theta(q,\mathbf{z}) \quad\text{with}\quad \mathbf{z}\sim p_\theta(\cdot\mid q).
\]
Each adjacent pair $\tau_t=(z_t\!\to\! z_{t+1})$ is a \emph{Wtransition} (claim, transformation, or sub-derivation).
Errors typically arise at a local transition $\tau_{t^\star}$ but remain hidden because $\mathbf{z}$ is not externalized.

\paragraph{Why naive retrieval misses the failing step.}
Standard retrieval conditions on the global query (either the input $q$ or the model’s draft answer $y$), yielding document sets
$\mathcal{D}^{(q)}=\text{TopK}\big(r_\phi(q)\big)$ and
$\mathcal{D}^{(y)}=\text{TopK}\big(r_\phi(y)\big)$
from a corpus $\mathcal{C}$ using retriever $r_\phi:\mathcal{X}\!\to\!2^{\mathcal{C}}$.
Define a \emph{step-alignment} functional $A:\!2^{\mathcal{C}}\!\times\!\mathcal{T}\!\to\![0,1]$ that measures whether a set of documents directly relevant or irrelevant a specific transition $\tau$ (e.g., contains the premise, rule, entity relation, or counterevidence needed for that step).
Empirically for the failing step $t^\star$,
\[
A\!\big(\mathcal{D}^{(q)},\tau_{t^\star}\big)\;\ll\;1
\quad\text{and}\quad
A\!\big(\mathcal{D}^{(y)},\tau_{t^\star}\big)\;\ll\;1,
\]
because relevance to $q$ or $y$ is topical, not \emph{step-indexed};
thus the wrong (or missing) micro-premise persists and the error propagates.

\paragraph{Objective.}
We seek a way to (i) elicit a provisional chain $\hat{\mathbf{z}}=\langle \hat z_1,\ldots,\hat z_T\rangle$ that reflects the model’s intended intermediate states for this instance, and (ii) retrieve by transition, not by $q$ or $y$, so that evidence is locally aligned:
\[
\begin{aligned}
\forall\, t\in\{1,\ldots,T{-}1\}:\quad
\mathcal{E}_t
&=\TopK\!\big(r_\phi(q,\hat\tau_t)\big),\\
\hat\tau_t
&=(\hat z_t\!\to\!\hat z_{t+1}).
\end{aligned}
\]
with the desideratum
$A(\mathcal{E}_t,\tau_t)\!\gg\!A(\mathcal{D}^{(q)},\tau_t),A(\mathcal{D}^{(y)},\tau_t)$.
Using step-aligned evidence $\mathcal{E}_{1:T-1}$, the model revises its reasoning to an updated chain
$\mathbf{z}'=\langle z'_1,\ldots,z'_T\rangle$ and emits
\[
y' \;=\; f_\theta\!\big(q,\mathbf{z}',\mathcal{E}_{1:T-1}\big),
\]
aiming for improved correctness and faithfulness.

\begin{algorithm*}[t]
\caption{\textsc{TARSE}: Test-Time Adaptation via Retrieval of Skills and Experience}
\label{alg:tarse}
\begin{algorithmic}[1]
\STATE \textbf{Input:} clinical case $x$, question $q$; base LLM $M_{\theta}$;
experience library $\mathcal{E}$; skills library $\mathcal{S}$;
retrievers $r_E, r_S$; top-$m$ experiences, top-$k$ skills;
TTT steps $n$, learning rate $\eta$; alignment threshold $\tau$.
\STATE \textbf{Output:} final answer $y$; (optional) verified chain $\mathbf{z}^\star$; evidence bundle $\{\widetilde{\mathcal{E}},\widetilde{\mathcal{S}}\}$.

\vspace{1mm}
\STATE \textbf{Step 1: Retrieve Experience (similar solved case)} \rcomment{Figure: Step 1}
\STATE $\mathcal{B} \leftarrow \TopM\!\big(r_E(q,x;\mathcal{E})\big)$
\rcomment{$\mathcal{B}=\{(q^{(i)},y^{(i)},\mathbf{z}^{(i)})\}_{i=1}^{m}$}

\vspace{1mm}
\STATE \textbf{Step 2: TARSE = lightweight TTT for chain alignment} \rcomment{Figure: Step 2}
\STATE $\theta_q \leftarrow \theta$ \rcomment{initialize query-adapted parameters}
\FOR{$j=1,\ldots,n$}
    \STATE $\mathcal{L}_{\text{exp}} \leftarrow \frac{1}{m}\sum_{i=1}^{m} -\log p_{\theta_q}\!\left(\mathbf{z}^{(i)} \mid q^{(i)},x^{(i)}\right)$
    \rcomment{align to retrieved experience traces}
    \STATE $\theta_q \leftarrow \theta_q - \eta \nabla_{\theta_q}\mathcal{L}_{\text{exp}}$
\ENDFOR

\vspace{1mm}
\STATE \textbf{Step 3: Hypothesis / Provisional chain + step-aware Skills retrieval} \rcomment{Figure: Step 3}
\STATE $\hat{\mathbf{z}}=\langle \hat z_1 \to \cdots \to \hat z_T\rangle \sim p_{\theta_q}(\cdot\mid q,x)$
\rcomment{experience-conditioned hypothesis generation}
\FOR{$t=1,\ldots,T-1$}
    \STATE $\hat{\tau}_t \leftarrow (\hat z_t \to \hat z_{t+1})$ \rcomment{current transition}
    \STATE $\mathcal{S}_t \leftarrow TopK\!\big(r_S(q,x,\hat{\tau}_t;\mathcal{S})\big)$ \rcomment{retrieve candidate skills}
    \STATE $\widetilde{\mathcal{S}}_t \leftarrow \{s\in\mathcal{S}_t:\mathrm{Align}(s;q,x,\hat{\tau}_t)\ge\tau\}$
    \rcomment{keep only step-aligned guideline/protocol rules}
\ENDFOR

\vspace{1mm}
\STATE \textbf{Step 4: Verify then answer */} \rcomment{Figure: Step 4}
\STATE $y \sim p_{\theta_q}(\cdot \mid q,x,\mathbf{z},\{\widetilde{\mathcal{S}}_t\}_{t=1}^{T-1})$
\rcomment{final answer follows verified therapeutic chain}
\STATE \textbf{return} $y$ (and $\mathbf{z}^\star=\mathbf{z}$).
\end{algorithmic}
\end{algorithm*}
\section{Experience \& Skills Library Construction}

\subsection{Skills Library: Guideline-Style Decision Rules}

We first initialize a large skills pool by collecting reusable clinical procedures from diverse guideline-style sources (e.g., CDC/NICE/WHO, textbooks, and other medical corpora). The sources and corpus statistics are summarized in Table~\ref{tab:combined_sources}.
Each skill item $s\in\mathcal{S}$ is normalized into an actionable rule format (e.g., \texttt{conditions $\rightarrow$ test $\rightarrow$ finding $\rightarrow$ action}) so it can be retrieved and applied as a step-level “gate” during reasoning.

We then construct an experience library from LogicalChain trajectories paired with QA. Each item is a triple $(q_i, y_i, \mathbf{z}_i)$, where $q_i$ is the question/case, $y_i$ is the verified answer, and $\mathbf{z}_i$ is the step-indexed logical chain.

\textbf{Human evaluation.} To validate the constructed experience items, we randomly sample 100 extracted instances and ask annotators to judge whether the LogicalChain is faithful and supports the paired QA. We find that 88\% of the sampled items are correct.

\subsection{Experience Library: Logical Chains \& QA Traces}

We curate LogicalChain trajectories to satisfy three principles: (i) decompose reasoning into atomic, verifiable transitions, (ii) keep steps grounded in entities present in the supporting context, and (iii) tightly couple the chain with QA so the chain is operational rather than decorative.

\paragraph{Design Principle 1 (Structured, Step-Indexed Reasoning).}
We represent an experience trace as a sequence of micro--inference states
$\mathbf{z}=\langle z_1 \!\to\! z_2 \!\to\! \cdots \!\to\! z_T\rangle$,
where each $z_t$ is a micro--inference state and each transition
$\tau_t=(z_t\!\to\! z_{t+1})$ captures one atomic step.
Let $\mathcal{A}$ denote a minimal set of atomic transition \emph{types}
(e.g., ``premise$\!\to$intermediate'', ``intermediate$\!\to$conclusion''),
and let $\mathcal{Z}$ be the set of valid traces.
Any valid trace $\mathbf{z}\in\mathcal{Z}$ can be decomposed into a finite sequence of atomic transitions
$\mathbf{z}=\langle \tau_{1},\tau_{2},\ldots,\tau_{T-1}\rangle$ with $\tau_{t}\in\mathcal{A}$.
This factorization makes experience (a) interpretable/verifiable at the transition level and (b) addressable by step, enabling step-aware retrieval and alignment.

As a concrete example in medicine, an atomic step might be “symptom $\rightarrow$ suspected drug class” or “drug class $\rightarrow$ therapeutic mechanism (rule application)”.
We build a large corpus of experience traces from sources that naturally contain solved-case reasoning (e.g., medical QA explanations, textbook solutions, and other verified rationales).
When only factual context is available (without an explicit trajectory), we synthesize a step-indexed chain under strict constraints (see Appendix~\ref{appendix:data}) and manually/automatically filter for faithfulness.

This yields an experience library
$\mathcal{E}=\{(q_i,y_i,\mathbf{z}_i)\}_{i=1}^{N}$,
where each $\mathbf{z}_i=\langle z_{i,1}\!\to\! \cdots \!\to\! z_{i,T_i}\rangle$ is a verified reasoning trajectory paired with $q_i$ and $y_i$.

\paragraph{Design Principle 2 (Entity-Grounded Abstraction).}
Let $\mathcal{U}$ be the universe of clinical entities, and let
$\phi(c)\subseteq \mathcal{U}$ extract entities mentioned in a case/context $c$.
For each experience item $(q_i,y_i,\mathbf{z}_i)$, we enforce entity alignment:
\[
\mathrm{Ent}(\mathbf{z}_i)\ \subseteq\ \phi(c_i),
\]
where $c_i$ is the supporting context for $q_i$ (e.g., explanation passage or case description).
This keeps each transition grounded and reduces spurious steps.

\paragraph{Design Principle 3 (QA-Coupled Trajectories).}
We require the answer to be entailed by the trajectory, i.e., $\mathbf{z}_i \models y_i$.
We also store step-level decompositions (e.g., $\tau_t$ as a retrievable key) so that, at test time, the model can retrieve experience by the \emph{missing step} rather than only by global similarity.

\begin{figure*}
    \centering
    \includegraphics[width=0.86\linewidth]{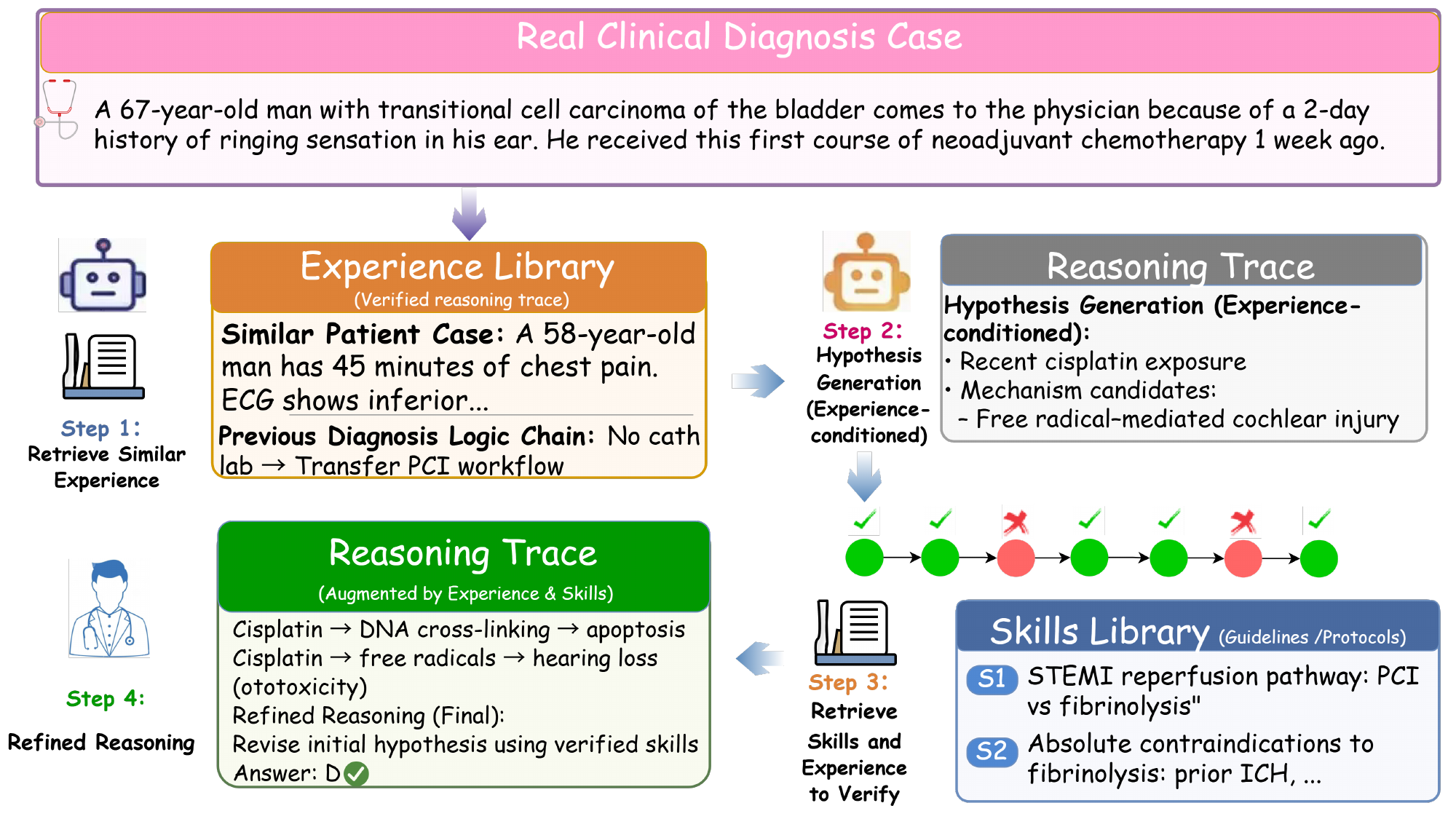}
    \caption{\textbf{TARSE process.}
       Starting from a real clinical diagnosis question (top), the agent first retrieves a similar verified experience trace from an Experience Library (Step 1), then performs experience-conditioned hypothesis generation to propose candidate mechanisms (Step 2). Next, it retrieves Skills from a Skills Library (e.g., guideline/protocol rules) together with supporting experience items to verify key transitions and enforce “gate” checks (Step 3), producing a refined reasoning trace that cleanly separates the side-effect chain (cisplatin → free radicals → ototoxicity/hearing loss) from the therapeutic mechanism chain (cisplatin → DNA cross-linking → apoptosis). The final answer follows the verified therapeutic chain (Step 4).}
    \label{fig:tarse}
    \vspace{-6mm}
\end{figure*}

\section{TARSE}


Humans handle clinical reasoning by combining experience and skills. Before answering, clinicians recall a similar solved case (experience) to sketch a tentative reasoning trajectory (an explicit step-by-step plan). 
They then consult guideline-style procedures (skills) to verify the fragile step (e.g., a contraindication gate, a mechanism-vs-side-effect disambiguation, or a required time window), thereby correcting the local error through step-aligned verification. The whole process is shown in Figure~\ref{fig:tarse}

We operationalize this workflow with \textsc{TARSE}. \textbf{Stage~A (experience alignment).} Given a test query, we retrieve a tiny minibatch of related solved cases from the \emph{experience library} (QA + verified reasoning traces) and perform lightweight test-time adaptation so the model commits to a provisional, instance-specific chain under the distribution of retrieved experience. \textbf{Stage~B (skills verification).} We then use each transition of the provisional chain as a step query to retrieve \emph{skills} (guideline/protocol rules) that are directly applicable to that step, and feed these rules back to the model to verify or repair the corresponding transition. 
This two-stage procedure mirrors human practice: first, ground the plan in prior experience; then, verify procedural correctness using skills within a single end-to-end inference loop.

\paragraph{Stage A: Test-time experience alignment and provisional chain.}
To externalize the model’s hidden reasoning for the current query, we perform a brief, neighbor-conditioned \emph{experience alignment} so the model first \emph{commits} to an explicit, step-indexed provisional chain before any skills are retrieved.
Given a test query $q$, a chain-aware retriever $r_{\phi}$ (with parameters $\phi$) returns a tiny exemplar set of chain triples 
$\mathcal{S}(q)=\{(q^{(i)},y^{(i)},\mathbf{z}^{(i)})\}_{i=1}^{m}$ 
of size $m$, where each $\mathbf{z}^{(i)}=\langle z^{(i)}_{1}\!\to\!\cdots\!\to\!z^{(i)}_{T_i}\rangle$ is a step-indexed logical chain paired with question $q^{(i)}$ and answer $y^{(i)}$. 
Starting from base model parameters $\theta$, we take a few gradient steps with learning rate $\eta_{\text{cot}}$ that \emph{only} align the output distribution $\pi_{\theta}(\cdot)$ to explicit chains:
\[
\theta_{q} \;=\; \theta \;-\; \eta_{\text{cot}}\,
\nabla_{\theta}\Bigg(\frac{1}{m}\sum_{i=1}^{m}\!-\log \pi_{\theta}\big(\mathbf{z}^{(i)} \mid q^{(i)}\big)\Bigg),
\]
yielding query-adapted parameters $\theta_{q}$. 
We then elicit a provisional chain for the instance:
\[
\begin{aligned}
\hat{\mathbf{z}}&=\langle \hat z_{1}\!\to\!\cdots\!\to\!\hat z_{T}\rangle \sim \pi_{\theta_{q}}(\cdot \mid q),\\
\hat{\tau}_{t}&=(\hat z_{t}\!\to\!\hat z_{t+1}),\quad t=1,\dots,T\!-\!1,
\end{aligned}
\]
where $T$ is the length of the elicited chain and $\hat{\tau}_{t}$ denotes its $t$-th transition. 
By collapsing diffuse latent hypotheses into a concrete sequence $\hat{\mathbf{z}}$, Stage~A makes intermediate states and transitions observable on this instance and furnishes the step queries $\{\hat{\tau}_t\}$ used in Stage~B.

\paragraph{Stage B: Step-aware skills retrieval and verification.}
To correct step-local mistakes, we retrieve \emph{skills} (guideline/protocol rules) conditioned on each transition of the provisional chain, rather than retrieving generic document chunks by the raw question.
Given the elicited chain $\hat{\mathbf{z}}$ and its transitions $\hat{\tau}_{t}$ for $t{=}1,\dots,T\!-\!1$, a skills retriever $r_{S}$ forms a step-conditioned query from $(q,\hat{\tau}_{t})$ and returns a small ranked set of candidate skill items:
\[
\mathcal{S}_{t} \;=\; \TopK_{k}\!\big(r_{S}(q,\hat{\tau}_{t})\big)
\quad\text{with}\quad
\mathcal{S}_{1:T}=\{\mathcal{S}_{t}\}_{t=1}^{T-1}.
\]
Optionally, a step-level alignment scorer $s_{\psi}(s\,;\,q,\hat{\tau}_{t})\!\in\![0,1]$ filters each pool by a threshold $\tau$:
\[
\widetilde{\mathcal{S}}_{t} \;=\; \{\, s \in \mathcal{S}_{t}\,:\, s_{\psi}(s\,;\,q,\hat{\tau}_{t}) \ge \tau \,\},
\qquad
\widetilde{\mathcal{S}}_{1:T}=\{\widetilde{\mathcal{S}}_{t}\}_{t=1}^{T-1}.
\]
We then condition the model on (i) the retrieved experience minibatch used in Stage~A and (ii) the step-aligned skills bundle, and generate the final answer (optionally emitting a verified/updated chain):
\[
y' \;\sim\; \pi_{\theta_{q}}\!\big(\,\cdot \,\big|\, q,\, \hat{\mathbf{z}},\, \widetilde{\mathcal{S}}_{1:T},\, \mathcal{B}\big).
\]
By conditioning retrieval on each transition $\hat{\tau}_{t}$ but retrieving \emph{skills as rules}, Stage~B supplies the exact gate/criterion/mechanism needed to validate or repair the fragile link in the chain, aligning the model’s stepwise plan with clinically valid procedures.

\begin{table*}[h]
\caption{Component usage, accuracy, and latency on MedQA. \emph{Time / Q} reports additional per-question latency relative to the Qwen-2.5 (CoT) baseline.}
\label{tab:method_cot_rag_ttt}
\centering
\begin{adjustbox}{width=0.75\textwidth}
\begin{tabular}{lcccc}
\toprule
 & \multicolumn{3}{c}{\textbf{Medical Domain}} & \multicolumn{1}{c}{\textbf{General Domain}} \\
\cmidrule(lr){2-4}\cmidrule(lr){5-5}
\textbf{Model} & \textbf{MedQA} & \textbf{MedMCQA} & \textbf{MMLU} & \textbf{MultiHopQA}\\
\midrule 
Qwen2.5-7B-Instruct  & 53.2 & 54.1 & 70.3 & 26.3  \\
Qwen2.5-14B-Instruct & 60.8 & 55.7 & 75.2 & 32.5  \\
MedRAG-Qwen2.5-7B    & 51.0 & 56.9 & 68.1 & 33.2  \\
i-MedRAG-Qwen2.5-7B  & 54.3 & 57.6 & 74.4 & 35.6  \\
rStar-Qwen2.5-7B     & 58.1 & 55.9 & 69.8 & 29.1 \\
rStar-Qwen2.5-14B    & 63.2 & 56.7 & 77.2 & 34.1 \\
TARSE-7B           & 70.1 & 62.1 & 75.4 & 45.1  \\
TARSE-14B          & \textbf{73.8} & \textbf{64.3} & \textbf{78.5} & \textbf{48.5}  \\
\bottomrule
\end{tabular}
\end{adjustbox}
\vspace{-3mm}
\end{table*}

\begin{table*}[t]
\caption{Model Performance split by \textit{Medical} vs \textit{General} domains.}
\label{tab:results}
\centering
\footnotesize
\setlength{\tabcolsep}{3pt}
\begin{adjustbox}{width=0.85\textwidth}
\begin{tabular}{lcccccc}
\toprule
\textbf{Method} & \textbf{Uses CoT} & \textbf{Uses RAG} & \textbf{Uses TTT} & \textbf{Accuracy (MedQA)} & \textbf{Time / Q (s)} & \textbf{Tokens} \\
\midrule
Qwen-2.5 (CoT)         & \cmark & \xmark & \xmark & 55.1 & +0s  & +0\\
Qwen-2.5-rStar         & \cmark & \xmark & \xmark & 58.1 & +497s & +491,132\\
Only TTT               & \xmark & \xmark & \cmark & 59.1 & +24s & +37,826\\
Only RAG-Skills (MedRAG)      & \xmark & \cmark & \xmark & 58.7 & +0s  & +2,641\\
CoT+RAG-Skills (iMedRAG)      & \cmark & \cmark & \xmark & 62.5 & +26s & +23,143\\
TTT+Experience                & \cmark & \xmark & \cmark & 59.3 & +25s & +39,177\\
TTT+Experience+RAG-Skills            & \cmark & \cmark & \cmark & 70.1 & +32s & +56,214\\
\bottomrule
\end{tabular}
\end{adjustbox}
\vspace{-3mm}
\end{table*}
\section{Experiment}
\subsection{Datasets \& Models}
\paragraph{Datasets}
We evaluate on three medical QA sets: MedQA~\citep{jin2021disease} (USMLE Step 1/2CK/3, Jun 2022–Mar 2023), MedMCQA~\citep{pal2022medmcqa} (194k MCQs across 21 subjects), and MMLU–Medical~\citep{hendrycks2020measuring} (nine medical areas), and two general multi-hop benchmarks: MultiHopQA~\citep{song2018graph}, which require cross-document evidence and step-wise verification.

\paragraph{Models}
Our backbone is Qwen2.5 (7B/14B)~\citep{hui2024qwen2}. Baselines include instruction-tuned LLMs (Qwen2.5-Instruct), RAG systems (MedRAG, i-MedRAG), and test-time reasoning (rStar–Qwen2.5); for broader comparison, we include open-source reasoning models and proprietary LLMs. TTT–RAG consistently outperforms open-source and RAG baselines across medical and general multi-hop tasks.

\subsection{Evaluation Settings}

To demonstrate the importance of retrieving the right experience together with the corresponding skills, we design experiments that directly compare our method against traditional inference and standard medical RAG baselines.

\paragraph{Reasoning trace quality (experience alignment).}
We evaluate whether retrieving and aligning to experience improves the quality of the model’s generated reasoning traces. Concretely, we prompt models to “think then answer” and compare a no-adaptation CoT baseline and strong open models (Qwen2.5 7B/14B/32B, Llama3.1 8B, Mistral-7B, Phi-4 14B) against our experience-aligned variant (TARSE-14B). We conduct both automatic and human evaluations. Automatically, we measure alignment between generated traces and gold references using string similarity and step-level entailment (\emph{Atomic Cov.}). For human evaluation, we randomly sample 100 traces from different model variants and ask annotators to rate them on: (1) logical coherence and (2) domain correctness. We report win rates against baselines.

\paragraph{Answer accuracy (skills-based verification and correction).}
We evaluate whether retrieving skills at the right step corrects errors and improves final accuracy. We use two complementary settings. Cross-domain comparison: on medical (MedQA, MedMCQA, MMLU–Medical) and general multi-hop suites (MultiHopQA, 2Wiki), we run the same backbones (Qwen2.5–7B/14B) and compare: (i) CoT baselines (Qwen2.5 CoT, rStar), (ii) question-level RAG (MedRAG), (iii) interactive CoT+RAG (i–MedRAG), and (iv) our method that retrieves experience to form a provisional chain and then retrieves step-aligned skills to verify/repair it before answering.

\textbf{Controlled ablations:} we further isolate components with three configurations matching backbone, prompts, and decoding: \textbf{Setting 1} TTT (no experience supervision, no skills retrieval; direct answer), \textbf{Setting 2} experience-aligned TTT (Logical) (with experience/trace supervision; still direct answer), and \textbf{Setting 3} TARSE (experience alignment + step-aware skills retrieval and verification). If gains appear primarily in Setting~3 and persist across domains against (i)–(iii), we attribute improvements to step-aligned skills verification rather than to longer reasoning or question-level retrieval.

\paragraph{Efficiency metrics.}
In addition to accuracy, we report efficiency in terms of both (i) additional wall-clock time per question and (ii) additional tokens per question, where token overhead is measured relative to the CoT baseline.

\section{Results}

\subsection{Main Results}
\paragraph{Reasoning trace quality.}
\begin{table}
    \caption{Performance of different models on reasoning quality in the MedQA dataset.}
    \label{tab:logical-chain-results}
    \centering
    \small
    \setlength{\tabcolsep}{3pt}
    \begin{adjustbox}{width=\columnwidth}
    \begin{tabular}{lccc}
    \toprule
    \textbf{Model} & \textbf{ROUGE-1} & \textbf{Atomic Cov.} & \textbf{Human} \\
    \midrule
    Qwen2.5 7B       & $0.2841\pm0.0063$ & $0.3541\pm0.0063$ & 31\%\\
    Llama3.1 8B       & $0.2640\pm0.0146$ & $0.3215\pm0.0270$ & 39\%\\
    Mistral-7B & $0.2617\pm0.0012$ & $0.3211\pm0.0021$ &30\%\\
    Phi-4 (14B)         & $0.2438\pm0.0012$ & $0.3213\pm0.0040$ & 45\%\\
    Qwen2.5 14B      & $0.2429\pm0.0646$ & $0.5437\pm0.0682$ & 41\%\\
    Qwen2.5 32B      & $0.2225\pm0.0003$  & $0.5936\pm0.0017$ & 41\%\\
    TARSE-14B       & $\textbf{0.3318}\pm0.0130$ & $\textbf{0.6112}\pm0.0210$ & --\\
    \bottomrule
    \end{tabular}
    \end{adjustbox}
    \vspace{-6mm}
\end{table}

We evaluate whether retrieving and aligning with experience (verified-reasoning traces from solved cases) improves the quality of the model’s generated reasoning trace for a new question. Compared with plain CoT baselines, lightweight test-time adaptation provides modest stabilization but still trails experience-aligned variants, as the trace-quality comparison on MedQA shows (Table~\ref{tab:logical-chain-results}). 
Among strong non-TTT open models, ROUGE-1 clusters in a narrow band (Qwen2.5-7B $0.2841\pm0.0063$, Llama3.1-8B $0.2640\pm0.0146$, Mistral-7B $0.2617\pm0.0012$, Phi-4-14B $0.2438\pm0.0012$), while step-level entailment (\emph{Atomic Cov.}) grows mainly with scale on the same backbone (Qwen2.5-14B $0.5437\pm0.0682$ $\rightarrow$ Qwen2.5-32B $0.5936\pm0.0017$), indicating larger models may assert more supported facts without necessarily producing step-aligned traces.
By contrast, our experience-aligned pipeline (TARSE-14B) achieves the strongest automatic trace quality, ROUGE-1 $0.3318\pm0.0130$ and \emph{Atomic Cov.} $0.6112\pm0.0210$, suggesting that retrieving similar solved-case experience and adapting to it helps the model externalize a more faithful, step-indexed reasoning trajectory.

\paragraph{Answer accuracy.}
Table~\ref{tab:results} summarizes performance split by \textit{medical} and \textit{general} domains. Across datasets, our method improves over strong baselines, suggesting that combining experience-grounded reasoning with step-aligned skill verification is effective for multi-hop and constraint-heavy problems.

\paragraph{Cross-domain pattern.}
A consistent trend is that TARSE yields larger gains on datasets where correctness depends on \emph{executing a specific procedural check at the right time} (e.g., exclusion criteria, contraindication gates, or mechanism-vs-side-effect disambiguation) rather than merely recalling a fact. This explains why TARSE improves not only in the medical domain (MedQA/MedMCQA/MMLU) but also transfers to a general multi-hop benchmark (MultiHopQA): the bottleneck is often step selection and verification, which is domain-agnostic.

\paragraph{Why question-level RAG is insufficient.}
Comparing MedRAG and i-MedRAG with TARSE indicates that retrieving more text does not reliably translate to higher accuracy when the retrieved evidence is not aligned to the failing transition. In Table~\ref{tab:method_cot_rag_ttt}, Only RAG (MedRAG) improves modestly over the CoT baseline but remains far below TTT+CoT+RAG. This gap supports our central claim: the dominant failure mode is not missing evidence per se, but \emph{misplaced evidence}---relevant snippets retrieved at the question level are often too coarse to correct a particular mistaken step.

\paragraph{Why TARSE can beat longer ``thinking''.}
Heavy test-time scaling (e.g., rStar) increases inference-time exploration, but it does not change what the model treats as the decision bottleneck. When errors originate from confusing two nearby causal chains (therapeutic mechanism vs. adverse-effect mechanism) or skipping a gate condition, sampling more chains can amplify the same wrong heuristic. In contrast, TARSE explicitly (i) anchors the provisional chain to similar verified experience and then (ii) injects step-aligned skills as executable rules to constrain the fragile link, which is a qualitatively different correction mechanism.

\paragraph{Practical takeaway.}
Overall, the results suggest a simple deployment guideline: if a benchmark (or real clinical setting) contains many questions whose label flips when a single intermediate check is applied or omitted, then adding \emph{step-aware skills verification} is more reliable than scaling up CoT length or retrieving more unstructured passages.

\subsection{Ablation Study.}

Table~\ref{tab:method_cot_rag_ttt} isolates the contributions of CoT, RAG, and test-time adaptation (TTT). Adding step-indexed supervision during TTT yields a modest but consistent gain when the model answers without retrieval (Only TTT: 59.1 vs.
Qwen-2.5 (CoT): 55.1; and TTT+CoT: 59.3), suggesting that aligning to experience helps the model produce cleaner intermediate premises.

The major correction appears when the reasoning chain is further \emph{verified with skills} via retrieval: TTT+CoT+RAG reaches 70.1 on MedQA, improving over CoT+RAG (iMedRAG: 62.5). This supports our hypothesis that step-aligned skill verification, rather than simply longer CoT, is the key driver of accuracy gains.

\subsection{Efficiency}
Table~\ref{tab:method_cot_rag_ttt} disentangles the accuracy\textendash latency\textendash token trade-off. Pure CoT (55.1, +0s) is the weakest. Adding more “thinking” without retrieval via rStar yields only a small gain (58.1, +497s), indicating poor efficiency. Either component alone helps modestly: Only RAG improves to 58.7 at no extra time (+3.6, +0s), while Only TTT reaches 59.1 with moderate overhead (+4.0, +24s). Combining CoT with generic interactive RAG (iMedRAG) produces a larger boost (62.5, +26s), showing synergy between reasoning and document access.

Besides latency, Table~\ref{tab:method_cot_rag_ttt} reports the additional token cost per question relative to CoT. Overall, the full TTT+CoT+RAG remains the best trade-off (70.1, +32s): it improves accuracy while keeping both latency and token overhead moderate compared with heavy test-time scaling baselines.

\section{Conclusion}

We present \textsc{LogicalChain + TARSE}, a retrieval-augmented reasoning framework that enhances LLM performance by integrating structured logical chains through test-time adaptation.
By dynamically aligning the model’s reasoning process with retrieved inference paths, TARSE significantly enhances performance on medical and mathematical benchmarks, surpassing existing retrieval-augmented and instruction-tuned baselines. Experimental results show that incorporating structured reasoning pathways during inference improves both the accuracy and logical coherence of model outputs. These findings demonstrate the effectiveness of logic-guided, instance-specific adaptation in advancing the reasoning capabilities of LLMs for complex tasks.

\section*{Impact Statement}

This work studies how to make a reasoning agent more reliable by retrieving two curated resources at test time: (1) clinical skills, which are reusable step-by-step rules from guidelines and protocols, and (2) clinical experience, which are verified reasoning traces from solved cases. The agent retrieves the right skill and the right experience for the current reasoning step, then uses lightweight test-time adaptation to keep its intermediate reasoning aligned with these resources and to avoid unsupported shortcuts. 
This can improve decision support for medical question answering, especially for multi-step questions that need explicit checks and branching. It can also generalize to other domains that have clear procedures and many past solved traces, such as safety checklists, debugging, compliance review, and scientific workflows. At the same time, medical deployment needs careful use. Wrong or outdated skills, biased experience traces, or wrong retrieval can still lead to wrong answers. Over-reliance on retrieved tool outputs can also mislead users. These systems should be used as decision support, not as autonomous diagnosis, and they should be tested with clinical safety protocols, audit logs for retrieved items, and strong privacy controls.

\bibliography{icml}

\begin{thebibliography}{46}
\providecommand{\natexlab}[1]{#1}
\providecommand{\url}[1]{\texttt{#1}}
\expandafter\ifx\csname urlstyle\endcsname\relax
  \providecommand{\doi}[1]{doi: #1}\else
  \providecommand{\doi}{doi: \begingroup \urlstyle{rm}\Url}\fi

\bibitem[Behrouz et~al.(2024)Behrouz, Zhong, and Mirrokni]{behrouz2024titans}
Behrouz, A., Zhong, P., and Mirrokni, V.
\newblock Titans: Learning to memorize at test time.
\newblock \emph{arXiv preprint arXiv:2501.00663}, 2024.

\bibitem[Cobbe et~al.(2021)Cobbe, Kosaraju, Bavarian, Chen, Jun, Kaiser, Plappert, Tworek, Hilton, Nakano, et~al.]{cobbe2021training}
Cobbe, K., Kosaraju, V., Bavarian, M., Chen, M., Jun, H., Kaiser, L., Plappert, M., Tworek, J., Hilton, J., Nakano, R., et~al.
\newblock Training verifiers to solve math word problems, 2021.
\newblock \emph{URL https://arxiv. org/abs/2110.14168}, 9, 2021.

\bibitem[Dziri et~al.(2023)Dziri, Lu, Sclar, Li, Jiang, Lin, Welleck, West, Bhagavatula, Le~Bras, et~al.]{dziri2023faith}
Dziri, N., Lu, X., Sclar, M., Li, X.~L., Jiang, L., Lin, B.~Y., Welleck, S., West, P., Bhagavatula, C., Le~Bras, R., et~al.
\newblock Faith and fate: Limits of transformers on compositionality.
\newblock \emph{Advances in Neural Information Processing Systems}, 36:\penalty0 70293--70332, 2023.

\bibitem[Guu et~al.(2020)Guu, Lee, Tung, Pasupat, and Chang]{guu2020retrieval}
Guu, K., Lee, K., Tung, Z., Pasupat, P., and Chang, M.
\newblock Retrieval augmented language model pre-training.
\newblock In \emph{International conference on machine learning}, pp.\  3929--3938. PMLR, 2020.

\bibitem[He et~al.(2024)He, Luo, Bai, Hu, Thai, Shen, Hu, Han, Huang, Zhang, et~al.]{he2024olympiadbench}
He, C., Luo, R., Bai, Y., Hu, S., Thai, Z.~L., Shen, J., Hu, J., Han, X., Huang, Y., Zhang, Y., et~al.
\newblock Olympiadbench: A challenging benchmark for promoting agi with olympiad-level bilingual multimodal scientific problems.
\newblock \emph{arXiv preprint arXiv:2402.14008}, 2024.

\bibitem[Hendrycks et~al.(2020)Hendrycks, Burns, Basart, Zou, Mazeika, Song, and Steinhardt]{hendrycks2020measuring}
Hendrycks, D., Burns, C., Basart, S., Zou, A., Mazeika, M., Song, D., and Steinhardt, J.
\newblock Measuring massive multitask language understanding.
\newblock \emph{arXiv preprint arXiv:2009.03300}, 2020.

\bibitem[Hu et~al.(2025)Hu, Zhao, Xu, Sun, Lou, Lin, Luo, and Rajmohan]{hu2025agentgen}
Hu, M., Zhao, P., Xu, C., Sun, Q., Lou, J.-G., Lin, Q., Luo, P., and Rajmohan, S.
\newblock Agentgen: Enhancing planning abilities for large language model based agent via environment and task generation.
\newblock In \emph{Proceedings of the 31st ACM SIGKDD Conference on Knowledge Discovery and Data Mining V. 1}, pp.\  496--507, 2025.

\bibitem[Hui et~al.(2024)Hui, Yang, Cui, Yang, Liu, Zhang, Liu, Zhang, Yu, Lu, et~al.]{hui2024qwen2}
Hui, B., Yang, J., Cui, Z., Yang, J., Liu, D., Zhang, L., Liu, T., Zhang, J., Yu, B., Lu, K., et~al.
\newblock Qwen2. 5-coder technical report.
\newblock \emph{arXiv preprint arXiv:2409.12186}, 2024.

\bibitem[Jiang et~al.(2025{\natexlab{a}})Jiang, Wang, Song, Hu, Zhou, Pu, Zhang, Yang, Feng, Zhou, et~al.]{jiang2025hulu}
Jiang, S., Wang, Y., Song, S., Hu, T., Zhou, C., Pu, B., Zhang, Y., Yang, Z., Feng, Y., Zhou, J.~T., et~al.
\newblock Hulu-med: A transparent generalist model towards holistic medical vision-language understanding.
\newblock \emph{arXiv preprint arXiv:2510.08668}, 2025{\natexlab{a}}.

\bibitem[Jiang et~al.(2025{\natexlab{b}})Jiang, Fang, Qiu, Zhang, Xu, Chen, Zhang, Zhang, Fang, Ma, et~al.]{jiang2025tc}
Jiang, X., Fang, Y., Qiu, R., Zhang, H., Xu, Y., Chen, H., Zhang, W., Zhang, R., Fang, Y., Ma, X., et~al.
\newblock Tc--rag: Turing--complete rag’s case study on medical llm systems.
\newblock In \emph{Proceedings of the 63rd Annual Meeting of the Association for Computational Linguistics (Volume 1: Long Papers)}, pp.\  11400--11426, 2025{\natexlab{b}}.

\bibitem[Jin et~al.(2025)Jin, Zeng, Yue, Yoon, Arik, Wang, Zamani, and Han]{jin2025search}
Jin, B., Zeng, H., Yue, Z., Yoon, J., Arik, S., Wang, D., Zamani, H., and Han, J.
\newblock Search-r1: Training llms to reason and leverage search engines with reinforcement learning.
\newblock \emph{arXiv preprint arXiv:2503.09516}, 2025.

\bibitem[Jin et~al.(2021)Jin, Pan, Oufattole, Weng, Fang, and Szolovits]{jin2021disease}
Jin, D., Pan, E., Oufattole, N., Weng, W.-H., Fang, H., and Szolovits, P.
\newblock What disease does this patient have? a large-scale open domain question answering dataset from medical exams.
\newblock \emph{Applied Sciences}, 11\penalty0 (14):\penalty0 6421, 2021.

\bibitem[Lee et~al.()Lee, Kim, Kang, Bang, Song, and Lee]{leera}
Lee, Y., Kim, D., Kang, J., Bang, J., Song, H., and Lee, J.-G.
\newblock Ra-tta: Retrieval-augmented test-time adaptation for vision-language models.
\newblock In \emph{The Thirteenth International Conference on Learning Representations}.

\bibitem[Lewis et~al.(2020)Lewis, Perez, Piktus, Petroni, Karpukhin, Goyal, K{\"u}ttler, Lewis, Yih, Rockt{\"a}schel, et~al.]{lewis2020retrieval}
Lewis, P., Perez, E., Piktus, A., Petroni, F., Karpukhin, V., Goyal, N., K{\"u}ttler, H., Lewis, M., Yih, W.-t., Rockt{\"a}schel, T., et~al.
\newblock Retrieval-augmented generation for knowledge-intensive nlp tasks.
\newblock \emph{Advances in neural information processing systems}, 33:\penalty0 9459--9474, 2020.

\bibitem[Li et~al.(2023)Li, Li, Zhang, Dan, Jiang, and Zhang]{li2023chatdoctor}
Li, Y., Li, Z., Zhang, K., Dan, R., Jiang, S., and Zhang, Y.
\newblock Chatdoctor: A medical chat model fine-tuned on a large language model meta-ai (llama) using medical domain knowledge.
\newblock \emph{Cureus}, 15\penalty0 (6), 2023.

\bibitem[Liang et~al.(2025)Liang, Zhang, Zhu, Wang, He, and Zhou]{liang2025rgar}
Liang, S., Zhang, L., Zhu, H., Wang, W., He, Y., and Zhou, D.
\newblock Rgar: Recurrence generation-augmented retrieval for factual-aware medical question answering.
\newblock \emph{arXiv preprint arXiv:2502.13361}, 2025.

\bibitem[Lightman et~al.(2023)Lightman, Kosaraju, Burda, Edwards, Baker, Lee, Leike, Schulman, Sutskever, and Cobbe]{lightman2023let}
Lightman, H., Kosaraju, V., Burda, Y., Edwards, H., Baker, B., Lee, T., Leike, J., Schulman, J., Sutskever, I., and Cobbe, K.
\newblock Let's verify step by step.
\newblock In \emph{The Twelfth International Conference on Learning Representations}, 2023.

\bibitem[Liu et~al.(2023)Liu, Ning, Teng, Liu, Zhou, and Zhang]{liu2023evaluating}
Liu, H., Ning, R., Teng, Z., Liu, J., Zhou, Q., and Zhang, Y.
\newblock Evaluating the logical reasoning ability of chatgpt and gpt-4.
\newblock \emph{arXiv preprint arXiv:2304.03439}, 2023.

\bibitem[Lozano et~al.(2023)Lozano, Fleming, Chiang, and Shah]{lozano2023clinfo}
Lozano, A., Fleming, S.~L., Chiang, C.-C., and Shah, N.
\newblock Clinfo. ai: An open-source retrieval-augmented large language model system for answering medical questions using scientific literature.
\newblock In \emph{Pacific Symposium on Biocomputing 2024}, pp.\  8--23. World Scientific, 2023.

\bibitem[Myers et~al.(2025)Myers, Dligach, Miller, Barr, Gao, Churpek, Mayampurath, and Afshar]{myers2025evaluating}
Myers, S., Dligach, D., Miller, T.~A., Barr, S., Gao, Y., Churpek, M., Mayampurath, A., and Afshar, M.
\newblock Evaluating retrieval-augmented generation vs. long-context input for clinical reasoning over ehrs.
\newblock \emph{arXiv preprint arXiv:2508.14817}, 2025.

\bibitem[Ou et~al.(2025)Ou, Huang, Zhao, Yu, Lu, and Ying]{ou2025experience}
Ou, J., Huang, T., Zhao, Y., Yu, Z., Lu, P., and Ying, R.
\newblock Experience retrieval-augmentation with electronic health records enables accurate discharge qa.
\newblock \emph{arXiv preprint arXiv:2503.17933}, 2025.

\bibitem[Pal et~al.(2022)Pal, Umapathi, and Sankarasubbu]{pal2022medmcqa}
Pal, A., Umapathi, L.~K., and Sankarasubbu, M.
\newblock Medmcqa: A large-scale multi-subject multi-choice dataset for medical domain question answering.
\newblock In \emph{Conference on Health, Inference, and Learning}, pp.\  248--260. PMLR, 2022.

\bibitem[Park(2025)]{park2025instruct}
Park, S.
\newblock Instruct-skillmix: A powerful pipeline for llm instruction tuning.
\newblock Master's thesis, Princeton University, 2025.

\bibitem[Sellergren et~al.(2025)Sellergren, Kazemzadeh, Jaroensri, Kiraly, Traverse, Kohlberger, Xu, Jamil, Hughes, Lau, et~al.]{sellergren2025medgemma}
Sellergren, A., Kazemzadeh, S., Jaroensri, T., Kiraly, A., Traverse, M., Kohlberger, T., Xu, S., Jamil, F., Hughes, C., Lau, C., et~al.
\newblock Medgemma technical report.
\newblock \emph{arXiv preprint arXiv:2507.05201}, 2025.

\bibitem[Song et~al.(2018)Song, Zhang, Wang, and Gildea]{song2018graph}
Song, L., Zhang, Y., Wang, Z., and Gildea, D.
\newblock A graph-to-sequence model for amr-to-text generation.
\newblock \emph{arXiv preprint arXiv:1805.02473}, 2018.

\bibitem[Sun et~al.(2020)Sun, Wang, Liu, Miller, Efros, and Hardt]{sun2020test}
Sun, Y., Wang, X., Liu, Z., Miller, J., Efros, A., and Hardt, M.
\newblock Test-time training with self-supervision for generalization under distribution shifts.
\newblock In \emph{International conference on machine learning}, pp.\  9229--9248. PMLR, 2020.

\bibitem[Team(2024)]{aimo2024amc}
Team, A.-M.
\newblock Ai-mo amc validation dataset.
\newblock \url{https://huggingface.co/datasets/AI-MO/aimo-validation-amc}, 2024.
\newblock Accessed: 2024-05-09.

\bibitem[Wang et~al.(2025{\natexlab{a}})Wang, Wang, Leong, and Li]{wang2025steca}
Wang, H., Wang, J., Leong, C.~T., and Li, W.
\newblock Steca: Step-level trajectory calibration for llm agent learning.
\newblock \emph{arXiv preprint arXiv:2502.14276}, 2025{\natexlab{a}}.

\bibitem[Wang et~al.(2024{\natexlab{a}})Wang, Yang, Yao, and Yu]{wang2024jmlr}
Wang, J., Yang, Z., Yao, Z., and Yu, H.
\newblock Jmlr: Joint medical llm and retrieval training for enhancing reasoning and professional question answering capability.
\newblock \emph{arXiv preprint arXiv:2402.17887}, 2024{\natexlab{a}}.

\bibitem[Wang et~al.(2024{\natexlab{b}})Wang, Yao, Yang, Zhou, Li, Wang, Xu, and Yu]{wang2024notechat}
Wang, J., Yao, Z., Yang, Z., Zhou, H., Li, R., Wang, X., Xu, Y., and Yu, H.
\newblock Notechat: a dataset of synthetic patient-physician conversations conditioned on clinical notes.
\newblock In \emph{Findings of the Association for Computational Linguistics: ACL 2024}, pp.\  15183--15201, 2024{\natexlab{b}}.

\bibitem[Wang et~al.(2022)Wang, Wei, Schuurmans, Le, Chi, Narang, Chowdhery, and Zhou]{wang2022self}
Wang, X., Wei, J., Schuurmans, D., Le, Q., Chi, E., Narang, S., Chowdhery, A., and Zhou, D.
\newblock Self-consistency improves chain of thought reasoning in language models.
\newblock \emph{arXiv preprint arXiv:2203.11171}, 2022.

\bibitem[Wang et~al.(2025{\natexlab{b}})Wang, Li, Chen, Zhu, Wu, Zhang, Xu, Chen, Fu, Wan, et~al.]{wang2025huatuo}
Wang, X., Li, J., Chen, S., Zhu, Y., Wu, X., Zhang, Z., Xu, X., Chen, J., Fu, J., Wan, X., et~al.
\newblock Huatuo-26m, a large-scale chinese medical qa dataset.
\newblock In \emph{Findings of the Association for Computational Linguistics: NAACL 2025}, pp.\  3828--3848, 2025{\natexlab{b}}.

\bibitem[Wang et~al.(2025{\natexlab{c}})Wang, Gandhi, Neubig, and Fried]{wang2025inducing}
Wang, Z.~Z., Gandhi, A., Neubig, G., and Fried, D.
\newblock Inducing programmatic skills for agentic tasks.
\newblock \emph{arXiv preprint arXiv:2504.06821}, 2025{\natexlab{c}}.

\bibitem[Wei et~al.(2022)Wei, Wang, Schuurmans, Bosma, Xia, Chi, Le, Zhou, et~al.]{wei2022chain}
Wei, J., Wang, X., Schuurmans, D., Bosma, M., Xia, F., Chi, E., Le, Q.~V., Zhou, D., et~al.
\newblock Chain-of-thought prompting elicits reasoning in large language models.
\newblock \emph{Advances in neural information processing systems}, 35:\penalty0 24824--24837, 2022.

\bibitem[Wu et~al.(2024{\natexlab{a}})Wu, Zhu, Qi, Chen, Xu, Menolascina, and Grau]{wu2024medical}
Wu, J., Zhu, J., Qi, Y., Chen, J., Xu, M., Menolascina, F., and Grau, V.
\newblock Medical graph rag: Towards safe medical large language model via graph retrieval-augmented generation.
\newblock \emph{arXiv preprint arXiv:2408.04187}, 2024{\natexlab{a}}.

\bibitem[Wu et~al.(2024{\natexlab{b}})Wu, Qiu, Ross, Aky{\"u}rek, Chen, Wang, Kim, Andreas, and Kim]{wu2024reasoning}
Wu, Z., Qiu, L., Ross, A., Aky{\"u}rek, E., Chen, B., Wang, B., Kim, N., Andreas, J., and Kim, Y.
\newblock Reasoning or reciting? exploring the capabilities and limitations of language models through counterfactual tasks.
\newblock In \emph{Proceedings of the 2024 Conference of the North American Chapter of the Association for Computational Linguistics: Human Language Technologies (Volume 1: Long Papers)}, pp.\  1819--1862, 2024{\natexlab{b}}.

\bibitem[Xia et~al.(2026)Xia, Chen, Wang, Liu, Zeng, Wang, Han, Zhou, Zhao, Chen, et~al.]{xia2026skillrl}
Xia, P., Chen, J., Wang, H., Liu, J., Zeng, K., Wang, Y., Han, S., Zhou, Y., Zhao, X., Chen, H., et~al.
\newblock Skillrl: Evolving agents via recursive skill-augmented reinforcement learning.
\newblock \emph{arXiv preprint arXiv:2602.08234}, 2026.

\bibitem[Xia et~al.(2025)Xia, Lin, Wu, Zhou, Li, Wan, Xing, Zheng, Li, Shan, et~al.]{xia2025medrek}
Xia, S., Lin, H., Wu, Y., Zhou, Y., Li, Z., Wan, Z., Xing, X., Zheng, Y., Li, X., Shan, C., et~al.
\newblock Medrek: Retrieval-based editing for medical llms with key-aware prompts.
\newblock \emph{arXiv preprint arXiv:2510.13500}, 2025.

\bibitem[Xiong et~al.(2024{\natexlab{a}})Xiong, Jin, Lu, and Zhang]{xiong2024benchmarking}
Xiong, G., Jin, Q., Lu, Z., and Zhang, A.
\newblock Benchmarking retrieval-augmented generation for medicine.
\newblock In \emph{Findings of the Association for Computational Linguistics ACL 2024}, pp.\  6233--6251, 2024{\natexlab{a}}.

\bibitem[Xiong et~al.(2024{\natexlab{b}})Xiong, Jin, Wang, Zhang, Lu, and Zhang]{xiong2024improving}
Xiong, G., Jin, Q., Wang, X., Zhang, M., Lu, Z., and Zhang, A.
\newblock Improving retrieval-augmented generation in medicine with iterative follow-up questions.
\newblock In \emph{Biocomputing 2025: Proceedings of the Pacific Symposium}, pp.\  199--214. World Scientific, 2024{\natexlab{b}}.

\bibitem[Xu et~al.(2025)Xu, Chan, Li, Aljunied, Yuan, Wang, Xiao, Chen, Liu, Li, et~al.]{xu2025lingshu}
Xu, W., Chan, H.~P., Li, L., Aljunied, M., Yuan, R., Wang, J., Xiao, C., Chen, G., Liu, C., Li, Z., et~al.
\newblock Lingshu: A generalist foundation model for unified multimodal medical understanding and reasoning.
\newblock \emph{arXiv preprint arXiv:2506.07044}, 2025.

\bibitem[Yan et~al.(2025)Yan, Yang, Huang, Nie, Ding, Li, Ma, Kersting, Pan, Sch{\"u}tze, et~al.]{yan2025memory}
Yan, S., Yang, X., Huang, Z., Nie, E., Ding, Z., Li, Z., Ma, X., Kersting, K., Pan, J.~Z., Sch{\"u}tze, H., et~al.
\newblock Memory-r1: Enhancing large language model agents to manage and utilize memories via reinforcement learning.
\newblock \emph{arXiv preprint arXiv:2508.19828}, 2025.

\bibitem[Ye et~al.(2023)Ye, Zhang, Wang, Xu, and Zhang]{ye2023natural}
Ye, R., Zhang, C., Wang, R., Xu, S., and Zhang, Y.
\newblock Natural language is all a graph needs.
\newblock \emph{arXiv preprint arXiv:2308.07134}, 2023.

\bibitem[Zakka et~al.(2024)Zakka, Shad, Chaurasia, Dalal, Kim, Moor, Fong, Phillips, Alexander, Ashley, et~al.]{zakka2024almanac}
Zakka, C., Shad, R., Chaurasia, A., Dalal, A.~R., Kim, J.~L., Moor, M., Fong, R., Phillips, C., Alexander, K., Ashley, E., et~al.
\newblock Almanac—retrieval-augmented language models for clinical medicine.
\newblock \emph{Nejm ai}, 1\penalty0 (2):\penalty0 AIoa2300068, 2024.

\bibitem[Zhang et~al.(2025)Zhang, Chen, Liu, Xue, Liao, Liu, Wang, Ning, Chen, Fu, et~al.]{zhang2025agent}
Zhang, K., Chen, X., Liu, B., Xue, T., Liao, Z., Liu, Z., Wang, X., Ning, Y., Chen, Z., Fu, X., et~al.
\newblock Agent learning via early experience.
\newblock \emph{arXiv preprint arXiv:2510.08558}, 2025.

\bibitem[Zhao et~al.(2025)Zhao, Liu, Yang, and Miao]{zhao2025medrag}
Zhao, X., Liu, S., Yang, S.-Y., and Miao, C.
\newblock Medrag: Enhancing retrieval-augmented generation with knowledge graph-elicited reasoning for healthcare copilot.
\newblock In \emph{Proceedings of the ACM on Web Conference 2025}, pp.\  4442--4457, 2025.

\end{thebibliography}
\bibliographystyle{icml2026}

\newpage
\appendix
\onecolumn
\section{Appendix}
\subsection{Robustness: Evaluation on Math Domain.}

For mathematical reasoning, we conduct a comprehensive evaluation on four widely-used benchmarks: MATH~\cite{lightman2023let}, AMC 2023~\citep{aimo2024amc}, Olympiad Bench~\cite{he2024olympiadbench}, and GSM8K~\cite{cobbe2021training}. We conduct further ablation studies to better understand the behavior of our TARSE framework:

\textbf{Impact of Small-Batch Size during Test-Time Training.} We systematically vary the size of the adaptation batch used in TTT (e.g., 1, 2, 4, 8 examples) to assess how the quantity of retrieved supervision affects the quality of adaptation and final accuracy. This ablation helps reveal the trade-off between adaptation speed and generalization performance.

\textbf{Parallel Multi-Query Test-Time Adaptation.} To simulate a multi-user inference setting and evaluate scalability, we group concurrent user queries (e.g., 3, 6, 9, 12 queries) and perform joint adaptation using a concatenated retrieval batch. This parallel strategy aims to amortize the cost of TTT while enabling more efficient resource sharing. We measure how this affects accuracy and adaptation stability across different levels of parallelism.

\paragraph{Parallelize Experiment.}
To investigate how different configurations of logical chain integration and test-time parallelization affect mathematical reasoning performance, we conduct two sets of ablation studies, visualized in Table~\ref{tab:ablation-math}.

\begin{figure}[!ht]
    \centering
    \includegraphics[width=0.4\linewidth]{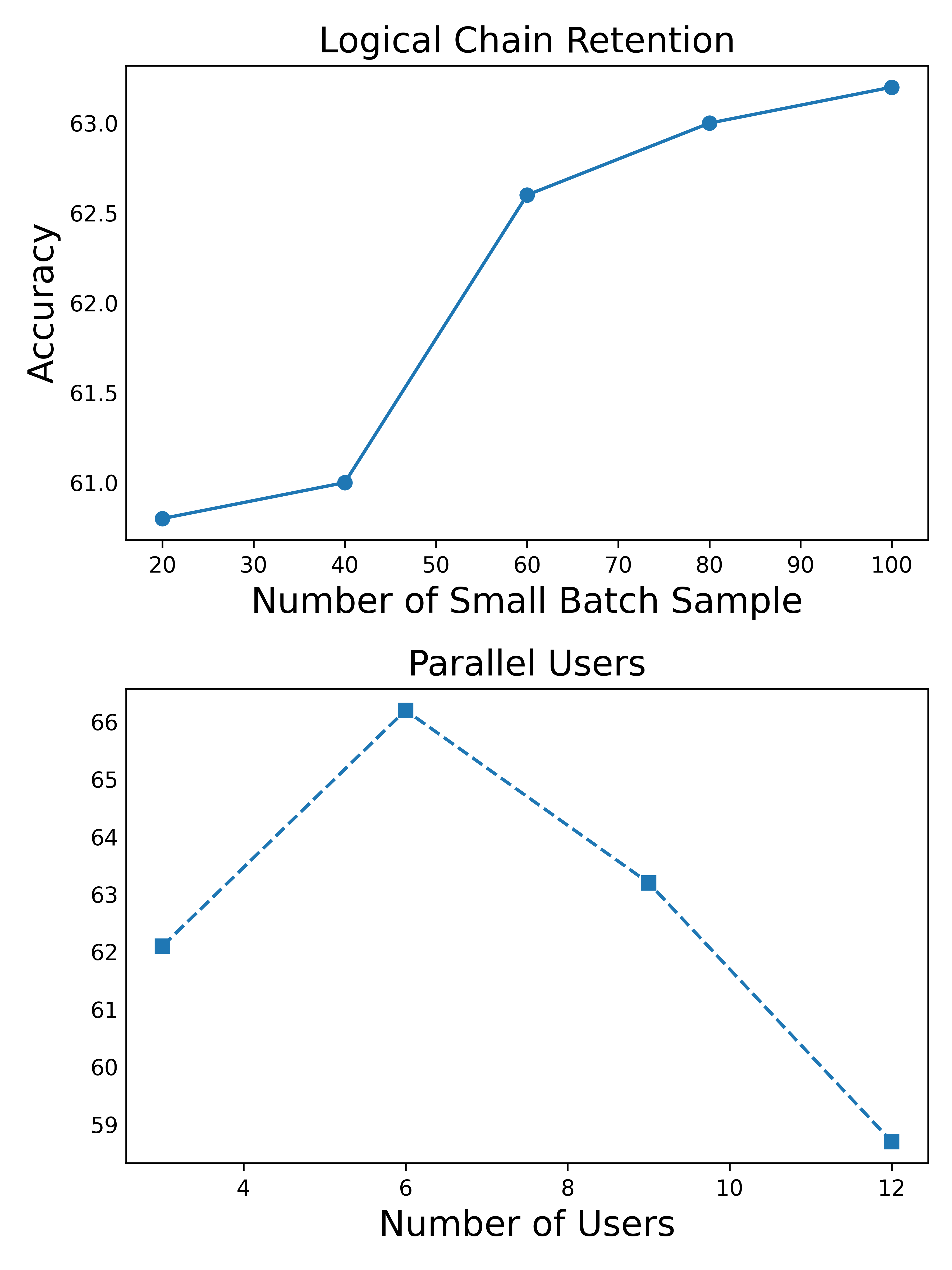}
    \caption{Ablation study for TARSE. (Top) Effect of the number of retained small-batch samples used in logical chain supervision. (Bottom) Impact of varying the number of parallel users during batch-wise test-time training.}
    \label{fig:ablation}
\end{figure}

In the Figure~\ref{fig:ablation}, we vary the \textbf{retention percentage} of retrieved logical chains during test-time training—from 20 to 100 samples and observe a monotonic increase in accuracy. This indicates that maintaining a higher proportion of logical-reasoning steps during adaptation substantially improves the model's ability to solve complex symbolic problems, increasing accuracy from 60.9\% to 63.2\%.
We also evaluate the effect of the parallel user batch size, in which retrieved batches from 3, 6, 9, or 12 users are aggregated and jointly used for test-time training. We find that moderate parallelism (6 users) improves performance by reducing variance in reasoning paths, whereas excessive parallelism (12 users) introduces noise and degrades accuracy. This suggests a trade-off between parallel efficiency and reasoning fidelity, and highlights the importance of controlled batch-level adaptation for symbolic reasoning tasks.


\begin{table}[!ht]
\caption{Comparison of model performance across mathematical benchmarks. rStar-Math shows strong performance across all metrics.}
\label{tab:ablation-math}
\centering
\scriptsize
\setlength{\tabcolsep}{3pt}
\begin{adjustbox}{max width=\columnwidth}
\begin{tabular}{lcccc}
\toprule
\textbf{Model}  & \textbf{MATH} & \textbf{AMC2023} & \textbf{Olympiad} & \textbf{GSM8K} \\
\midrule
Qwen2-7B-Instruct &  49.6  & 25.0 & 0.04 & 82.3 \\
NuminaMath-7B-COT & 55.8  & 27.5  &0.03 & 76.3  \\
rStar-Qwen2 & 60.4  & 30.0 & 20.0 & \textbf{87.2} \\
\textbf{TARSEs} & \textbf{63.2}  & \textbf{30.0} & \textbf{24.7} & 87.0 \\
\bottomrule
\end{tabular}
\end{adjustbox}

\end{table}
\subsection{Error Analysis}

To better understand the limitations of different retrieval strategies, we conducted an error analysis comparing a baseline RAG approach with our proposed TARSE framework. Using Qwen2.5-14B as the base model, we retrieve documents and directly generate answers, and compared this with TARSE’s test-time adapted retrieval and inference. We randomly sampled 100 test cases and manually categorized each incorrect prediction into one or more of the following four error types: (1) irrelevant or missing evidence, (2) misleading evidence, (3) insufficient reasoning ability, and (4) forgotten knowledge.

The ColBERT-RAG setup produced 15 errors due to irrelevant or missing documents, 6 from misleading content, 20 due to reasoning failures, and 6 involving forgotten background knowledge. In contrast, TARSE reduced these numbers to 7, 0, 12, and 5 respectively. These results highlight that TARSE not only improves retrieval relevance but also reduces reasoning and knowledge-related errors, suggesting that test-time adaptation provides more reliable and contextually aligned inputs for downstream inference.

\subsection{Data Construction Pipeline}
\label{appendix:data}

We categorize documents into two types:  
(1) \textbf{Expert-authored documents with explicit reasoning traces}, such as clinical guidelines, medical board exam explanations, or textbook problem solutions. These documents often contain structured logic that can be extracted or lightly reformatted to form usable reasoning chains.  
(2) \textbf{Factual documents without explicit reasoning}, including textbook paragraphs, curated QA datasets, and encyclopedic resources such as Wikipedia. These texts provide rich medical knowledge but require additional processing to reveal the underlying reasoning structure. The details are shown in Table~\ref{tab:prompt_part1}

For type (2), we synthesize logical chains by prompting GPT-4 with a structured instruction template. Each input includes a factual medical passage, and the model is asked to generate a step-by-step logical chain that connects context to conclusion while avoiding large logical jumps.  
We include constraints to ensure that:
\begin{itemize}
    \item The logical chain begins with the key facts or conditions in the document;
    \item Intermediate reasoning steps are included (e.g., physiological mechanisms, diagnostic steps, or treatment justifications);
    \item The final step logically derives the medically relevant conclusion (e.g., diagnosis, mechanism, treatment);
    \item Language remains generalizable and concise, avoiding unnecessary specifics (e.g., age, lab values) unless clinically necessary.
\end{itemize}

A representative prompt looks like this:

\begin{quote}
\texttt{“The text below contains a medical fact. Your task is to generate a logical chain that explains the underlying reasoning. Be concise but complete. Return format: Logical Chain: A $\rightarrow$ B $\rightarrow$ C”}
\end{quote}

This process allows us to convert static medical knowledge into dynamic reasoning trajectories that can be retrieved and aligned at inference time.
\begin{table*}[h!]
\centering
\caption{Prompt Template for Logical Rule Generation.}
\label{tab:prompt_part1}
\begin{adjustbox}{width=\textwidth}
\begin{tabular}{>{\centering\arraybackslash}m{4cm}p{12cm}}
\toprule
\rowcolor{gray!20}\textbf{System Prompt} & The text below contains medical information. Your task is to create rules based on the medical knowledge presented, ensuring each rule comprehensively represents the information in the evidence text without omitting any significant details. \\
\midrule
\textbf{1. Clear Distinction Between Conditions and Outcomes} & Each rule should clearly reflect when certain items are conditions, patient characteristics, or prerequisites (such as specific diagnoses, clinical findings, or circumstances) and when others are outcomes, actions, or recommendations. Always list conditions at the start of the rule, followed by the recommended actions or results. \\
\rowcolor{gray!20}\textbf{2. Multiple Rules for Complex Evidence} & If an evidence statement includes multiple distinct aspects, create separate rules for each part. This ensures that each concept is represented fully and clearly, preventing unrelated ideas from being merged into a single rule. Reuse the evidence as needed for each distinct rule. \\
\textbf{3. Parallel Items and Logical Structure} & Arrange items logically within each rule. 
\begin{itemize}
    \item For items that have a natural sequence based on real-world logic (e.g., patient characteristics $\rightarrow$ diagnostics $\rightarrow$ treatment recommendations), follow this order.
    \item If items are parallel without a strict sequence, indicate them at the same level within the rule without imposing order (e.g., “PET benefits $\rightarrow$ Diagnosis, Staging, Treatment Response”).
\end{itemize} \\
\rowcolor{gray!20}\textbf{4. Real-World Logical Order} & When applicable, follow a natural sequence based on real-world relationships rather than sentence order. The preferred structure is:
\begin{itemize}
    \item Conditions or Requirements
    \item Diagnostics
    \item Findings or Results
    \item Treatment Recommendations or Actions
    \item Follow-Up or Monitoring Plans (if applicable)
\end{itemize}
Present these elements in this logical order, regardless of how they are arranged in the original text. \\
\textbf{5. Reconstruct Guideline} & Ensure that the generated set of evidence and rules covers the guideline comprehensively, allowing the guideline to be reconstructed accurately from these parts. Use exact phrasing from the original guideline text where feasible. \\
\textbf{6. Include Every Key Detail} & Represent every significant term, phrase, or concept from the evidence. Each rule should be as concise as possible while fully representing all critical details. Avoid redundancy, but include all essential information. \\
\rowcolor{gray!20}\textbf{7. Direct Language Without Conditionals} & Use clear and direct language without excessive use of conditionals like “if” or “when.” Each rule should be logically complete and self-contained, conveying the full context of the information. \\
\bottomrule
\end{tabular}
\end{adjustbox}
\end{table*}

\clearpage

\begin{table*}[t]
\centering
\begin{adjustbox}{width=\textwidth}
\begin{tabular}{>{\centering\arraybackslash}m{4cm}p{12cm}}
\textbf{8. Output Format} & 
\textbf{Key Terms to Include}: \texttt{\{key\_words\}} \\
\textbf{Medical Text}: \texttt{\{medical\_text\}} \\
\vspace{1mm}

\textbf{Example Rule Format:} \\

\rowcolor{gray!10} \textbf{} & \textbf{Correct:} \\
\textbf{} & \textbf{Evidence}: “A patient presents with abdominal pain, followed by a CT scan, which shows an abscess. Treatment includes antibiotics.” \\
\textbf{} & \textbf{Rule}: Abdominal pain $\rightarrow$ CT scan $\rightarrow$ Abscess detected $\rightarrow$ Treatment: Antibiotics. \\
\rowcolor{gray!10} \textbf{} & \textbf{Incorrect:} \\
\textbf{} & \textbf{Evidence}: “A patient presents with abdominal pain, followed by a CT scan, which shows an abscess. Treatment includes antibiotics.” \\
\textbf{} & \textbf{Rule}: Abdominal pain $\rightarrow$ Abscess detected $\rightarrow$ CT scan $\rightarrow$ Treatment: Antibiotics. \\
\midrule
\textbf{Return Format} & 
\begin{itemize}
    \item Evidence:
    \item Rule:
    \item Evidence:
    \item Rule:
\end{itemize}
Return only the specified output format without additional commentary or explanation. \\
\bottomrule
\end{tabular}
\end{adjustbox}
\end{table*}

\subsection{Retriever Training and Evaluation}
\subsubsection{Logical Chain Retriever Training}

We retrieve logical chains using a two-stage retrieval–reranking pipeline~\cite{}. First, we use ColBERT to retrieve the top-K candidate logical chains for each query $q$. Then, a cross-encoder is applied to rerank these candidates. This pipeline balances retrieval efficiency with reranking effectiveness, leveraging ColBERT’s fast dense retrieval and the cross-encoder’s stronger semantic matching capabilities. Specifically, the cross encoder is trained to map a query \( q \) (typically a user question) to its corresponding logical chain \( l^+ \), sampled from a shared document \( d \). Each training instance is constructed as a triplet \((q, l^+, \{l^-_i\}_{i=1}^{N})\), where \(l^+\) is the gold reasoning path and \(l^-_i\) are hard negatives sampled from neighboring chains within the same document \(d\). The training objective minimizes the following contrastive loss:

\begin{equation}
\mathcal{L} = -\log \frac{\exp(\text{Sim}(q, l^+))}{\exp(\text{Sim}(q, l^+)) + \sum_{i=1}^{N} \exp(\text{Sim}(q, l^-_i))},
\end{equation}

where \( \text{Sim}(q, l) \) denotes the similarity score between query \(q\) and candidate logical chain \(l\), computed via a dual-encoder architecture. The softmax denominator encourages the model to assign the highest score to the gold chain.

Each query \(q\) is associated with a unique logical chain \(l\), such that the retrieved logical chain \(l^+\) implies both the supporting document and the reasoning steps required to answer \(q\). This alignment enables the retriever to support dual downstream tasks. QA-to-Chain Retrieval: Given a clinical question \(q\), retrieve the corresponding logical chain \(l^+\) that reflects guideline-based reasoning. Guideline-to-Chain Retrieval: Given a document excerpt \(d\), retrieve an appropriate reasoning path \(l^+\) for diagnostic or therapeutic inference. By training jointly on both tasks, the retriever learns generalizable alignment between questions, reasoning chains, and medical documents, supporting structured retrieval across diverse input scenarios.
\subsubsection{Evaluaton Results}
\label{appendix:retriever}
We compare the retrieval accuracy of our proposed Logical Chain Retriever against a strong baseline, ColBERT Retriever, on top-$k$ evidence selection. For the test datasets, we construct QA pairs based on Pubmed corpus to test. As shown in Table~\ref{tab:retriever_comparison}, the Logical Chain Retriever consistently outperforms ColBERT across all values of $k$. Notably, it achieves 73.27\% top-5 accuracy, compared to 62.33\% with ColBERT. This performance gap widens with larger $k$, reaching 81.79\% at top-20 retrieval, demonstrating the effectiveness of incorporating structured reasoning signals during retrieval. These results indicate that explicitly modeling logical chains not only improves alignment with the query intent, but also facilitates retrieval of more relevant and explanatory evidence.

\begin{table}
\caption{Top-$k$ retrieval accuracy (\%) comparison between Logical Chain Retriever and ColBERT Retriever.}
\label{tab:retriever_comparison}
\centering
\small 
\setlength{\tabcolsep}{4pt} 
\renewcommand{\arraystretch}{1.1} 
\begin{tabular}{lcccc}
\toprule
Logical Chain Retriever & 73.27 & 77.49 & 80.13 & 81.79 \\
ColBERT Retriever       & 62.33 & 65.12 & 67.48 & 70.51 \\
\bottomrule
\end{tabular}
\end{table}

\subsection{QA Generation Examples}
\label{appendix:qaexamples}

Given a logical chain, we automatically generate QA pairs by framing the final step as the answer and rephrasing the earlier reasoning path as a question. This alignment ensures that the question is logically entailed by the steps in the chain, and that the answer is the natural conclusion of the reasoning process.

For example, given the input:

\textbf{Document}: “Cisplatin binds DNA and induces cross-linking, which causes apoptosis.”\\
\textbf{Logical chain}: “Cisplatin $\rightarrow$ DNA crosslinking $\rightarrow$ blocks replication $\rightarrow$ apoptosis”

We generate:

\textbf{Q}: What is the therapeutic mechanism of Cisplatin?\\
\textbf{A}: DNA cross-linking

This question-answer pair reflects both the factual content and the reasoning structure, enabling it to be used in alignment tasks or as a target for supervised fine-tuning. When needed, we apply controlled variation (e.g., passive voice, question style) to increase linguistic diversity while preserving the underlying logic.

\subsection{Evaluation Prompt Design}
\label{template}
When evaluating close QA, we only need to calculate accuracy. However, many open QA tasks, such as diagnostic reasoning questions in the Amboss Dataset, present additional challenges. Although several methods exist for measuring textual similarity, such as F1 or ROUGE, both approaches have significant limitations in the medical domain. Therefore, we propose a very strict evaluation pipeine by using two evaluation metrics: the USMLE-Factuality score and the GPT-4o score. For the GPT-4o score, directly allowing GPT-4o to grade the answers is often ineffective, as GPT-4o tends to favor answers that align with its preferred linguistic style, which may not match our intended criteria. Thus, we introduce a scoring framework to evaluate model's fine grained diagnostic ability based on three aspects: \textbf{Key Points}, \textbf{Inference}, and \textbf{Evidence} which is designed by doctors:
\begin{itemize}
    \item \textbf{Key Points} assess whether the model's answer includes the critical elements present in the ground truth.
    \item \textbf{Inference} evaluates whether the diagnostic reasoning in the model's answer is correct, follows the same steps as the ground truth, and whether any key steps are omitted.
    \item \textbf{Evidence} examines whether the model's answer provides the crucial evidence to support its conclusions or diagnostic reasoning.
\end{itemize}

Finally, an average score will be calculated to represent the overall quality of the answer. To further reduce the influence of linguistic style on GPT-4's scoring, we propose revising all model-generated answers through GPT-4, ensuring that all outputs align with GPT-4's own style distribution. During this revision, GPT-4 will only see the model's answer, without access to any other information.

When scoring, GPT-4 will generate its own summaries of \textbf{Key Points}, \textbf{Inference}, and \textbf{Evidence} based on the ground truth. When assigning scores to these aspects, GPT-4 will no longer see the original answer but will only reference its summarized \textbf{Key Points}, \textbf{Inference}, and \textbf{Evidence}.

\label{prompt}
\begin{table*}[h!]
\centering
\caption{Evaluation Pipeline Prompt Example Template.}
\begin{adjustbox}{width=\textwidth}
\begin{tabular}{>{\centering\arraybackslash}m{3cm}p{5cm}p{5cm}} 
\rowcolor{gray!20}\multirow{3}{*}{\parbox[c][\height]{3cm}{\centering \textbf{System Prompt}}} & \multicolumn{2}{p{10cm}}{The text below contains medical information. Your task is to create rules based on the medical knowledge presented, ensuring each rule comprehensively represents the information in the evidence text without omitting any significant details.
} \\

\multirow{4}{*}{\textbf{Clear Distinction}} & \multicolumn{2}{p{10cm}}{Each rule should clearly reflect when certain items are conditions, patient characteristics, or prerequisites (such as specific diagnoses, clinical findings, or circumstances) and when others are outcomes, actions, or recommendations. Always list conditions at the start of the rule, followed by the recommended actions or results.} \\
\rowcolor{gray!20}\multirow{3}{*}{\parbox[c][\height]{3cm}{\centering \textbf{Multiple Rules for Complex Evidence}}} & \multicolumn{2}{p{10cm}}{If an evidence statement includes multiple distinct aspects, create separate rules for each part. This ensures that each concept is represented fully and clearly, preventing unrelated ideas from being merged into a single rule. Reuse the evidence as needed for each distinct rule.} \\
\multirow{5}{*}{\parbox[c][\height]{2.5cm}{\centering \textbf{Parallel Items and Logical Structure}}} & \multicolumn{2}{p{10cm}}{Arrange items logically within each rule.
   - For items that have a natural sequence based on real-world logic (e.g., patient characteristics -> diagnostics -> treatment recommendations), follow this order.
   - If items are parallel without a strict sequence, indicate them at the same level within the rule without imposing order, such as "PET benefits -> Diagnosis, Staging, Treatment Response" if they all apply equally.
} \\
\rowcolor{gray!20}\multirow{7}{*}{\parbox[c][\height]{3cm}{\centering \textbf{Real-World Logical Order}}} & \multicolumn{2}{p{10cm}}{Over Sentence Structure**: When applicable, follow a natural sequence based on real-world relationships rather than sentence order. The preferred sequence generally follows:
   - **Conditions or Requirements** (e.g., patient characteristics or specific criteria)
   - **Diagnostics** (e.g., tests performed)
   - **Findings or Results** (e.g., outcomes of diagnostics)
   - **Treatment Recommendations or Actions** (e.g., proposed treatments based on findings)
   - **Follow-Up or Monitoring Plans** (if applicable) 
   Present these elements in this logical order, regardless of how they are arranged in the original text.} \\
\multirow{3}{*}{\parbox[c][\height]{2.5cm}{\centering \textbf{Reconstruct Guideline}}} & \multicolumn{2}{p{10cm}}{Ensure that the generated set of evidence and rules covers the guideline comprehensively, allowing the guideline to be reconstructed accurately from these parts. Each evidence should be presented as close to the original guideline text as possible, using exact phrasing where feasible.} \\
\rowcolor{gray!20}\multirow{5}{*}{\parbox[c][\height]{3cm}{\centering \textbf{Include Details}}} & \multicolumn{2}{p{10cm}}{Represent every significant term, phrase, or concept from the evidence. Each rule item should be as concise as possible while fully representing all critical details. Avoid unnecessary redundancy but ensure all essential information is included.
Use clear and direct language without excessive use of conditionals like “if” or “when.” Each rule should be logically complete and self-contained, conveying the full context of the information.
} \\
\bottomrule
\end{tabular}
\end{adjustbox}
\label{evaluate_prompt1}
\end{table*}

\clearpage

\begin{table*}[t]
\centering
\caption{Evaluation Pipeline Prompt Example Template (continued).}
\begin{adjustbox}{width=\textwidth}
\begin{tabular}{>{\centering\arraybackslash}m{3cm}p{5cm}p{5cm}} 
\multirow{8}{*}{\parbox[c][\height]{3cm}{\centering \textbf{Output Format}}} & \multicolumn{2}{p{10cm}}{
**Key Terms to Include**: key words
**Medical Text**: medical text
**Medical Path**: medical path
**Example Rule Format**:
- **Correct**:
   - **Evidence**: "A patient presents with abdominal pain, followed by a CT scan, which shows an abscess. Treatment includes antibiotics."
   - **Rule**: Abdominal pain -> CT scan -> Abscess detected -> Treatment: Antibiotics.
- **Incorrect** (based purely on sentence order):
   - **Evidence**: "A patient presents with abdominal pain, followed by a CT scan, which shows an abscess. Treatment includes antibiotics."
   - **Rule**: Abdominal pain -> Abscess detected -> CT scan -> Treatment: Antibiotics.
**Return Format**:
1. Evidence:
   Rule:
2. Evidence:
   Rule:
Return only the specified output format without additional commentary or explanation.} \\
\bottomrule
\end{tabular}
\end{adjustbox}
\end{table*}

\begin{table*}[h!]
\centering
\caption{Evaluation Pipeline Prompt Example Template.}
\begin{tabular}{>{\centering\arraybackslash}m{3cm}p{6cm}p{6cm}} 
\rowcolor{gray!20}\multirow{4}{*}{\parbox[c][\height]{3cm}{\centering \textbf{Extract Key Points}}} & \multicolumn{2}{p{10cm}}{Based on the question and answer, summarize ten key points that you consider to be the most crucial from the standard answer. Return the response in the following format: \{1.2.3....\}
Here is the question:\{question\}
Here is the answer:\{answer\}
Please do not provide any additional information.} \\

\multirow{3}{*}{\textbf{Key Points}} & \multicolumn{2}{p{10cm}}{1. Multifocal electroretinogram (ERG) showed reduced signal in the right eye throughout the macula, confirming the diagnosis of AZOOR.2. Acute zonal occult outer retinopathy (AZOOR) was first described by Gass in 1993...} \\
\rowcolor{gray!20}\multirow{4}{*}{\parbox[c][\height]{3cm}{\centering \textbf{Extract Diagnostic Reasoning}}} & \multicolumn{2}{p{10cm}}{Based on the question and answer, please provide a detailed summary of the diagnostic reasoning from the standard answer. Return the response in the following format: \{1.2.3....\}
Here is the question:\{question\}
Here is the answer:\{answer\}
Please do not provide any additional information.} \\
\multirow{4}{*}{\parbox[c][\height]{2.5cm}{\centering \textbf{Diagnostic Reasoning}}} & \multicolumn{2}{p{10cm}}{1. The patient is a 7-year-old boy with a slowly growing, asymptomatic lump on the left lower neck since birth.2. Physical examination showed a yellowish, hump-like mass with a hairy surface and cartilage-like consistency near the left sternocleidomastoid muscle...} \\
\rowcolor{gray!20}\multirow{4}{*}{\parbox[c][\height]{3cm}{\centering \textbf{Extract Evidence}}} & \multicolumn{2}{p{10cm}}{Based on the question and answer, please provide a detailed evidence list which is proposed by correct answer. Return the response in the following format:  \{1.2.3....\}
Here is the question:\{question\}
Here is the answer:\{answer\}
Please do not provide any additional information.} \\
\multirow{5}{*}{\parbox[c][\height]{2.5cm}{\centering \textbf{Evidence}}} & \multicolumn{2}{p{10cm}}{1. Slowly growing, asymptomatic lump on left lower neck since birth.2. Physical examination revealed a yellowish, hump-like mass with hairy surface and cartilage-like consistency.3. Ultrasonography indicated a hypoechoic, avascular, bulging nodule with an anechoic tubular structure.4. MRI demonstrated a protuberant nodule with diffuse...} \\
\rowcolor{gray!20}\multirow{14}{*}{\parbox[c][\height]{3cm}{\centering \textbf{Key Points Score}}} & \multicolumn{2}{p{10cm}}{Act as a USMLE evaluator, your role involves assessing and comparing a medical student's explanation to the provided target answer. Begin the assessment by carefully reviewing the provided target answer. Then, based on following specific criteria, determine the score for the student's answer.
Please judge whether medical student's answer include these key points(or some other relevant points. But the amount of points must be complete). For example, ground truth have 10 key points, if student answer include one key he will get 0.5 point(if the answer include 5 points so should be 2.5).
Medical student's answer: 
\{answer\}
Key Points:
\{Key Point\}
Please only return a float number(from 0 to 5). You should check each point one by one(shouldn't judge based on language style such as fluence and so on. Only judge based on whether the student's answer include correct or relevant and complete key points). Don't generate any other information.
} \\
\end{tabular}
\label{evaluate_prompt2}
\end{table*}

\begin{table*}[h!]
    \centering
    \scriptsize
    \setlength{\tabcolsep}{3pt}
    \caption{Sources of Medical Guidelines, Information, and Corpus Statistics}
    \begin{adjustbox}{max width=\textwidth}
    \begin{tabular}{llcccccc}
        \toprule
        \textbf{Source/Corpus} & \textbf{Full Name} & \textbf{Guidelines/\#Doc.} & \textbf{Words/\#Snippets}  & \textbf{Audience/Domain} & \textbf{Country} & \textbf{Released} \\
        \midrule
        CCO & Cancer Care Ontario  & 87 & 199K  & Doctor & Canada & Yes \\
        CDC & Center for Disease Control and Prevention  & 621 & 6.7M  & Doctor & USA & Yes \\
        CMA & Canadian Medical Association  & 431 & 1.7M  & Doctor & Canada & Yes \\
        ICRC & International Committee of the Red Cross  & 49 & 1.2M & Doctor & International & Yes \\
        NICE & National Institute for Health and Care Excellence  & 1656 & 8.1M  & Doctor & UK & Yes \\
        PubMed & PubMed  & 1627 & 10.8M  & Doctor & Mix & Yes \\
        SPOR & Strategy for Patient-Oriented Research  & 217 & 1.1M & Doctor & Canada & Yes \\
        WHO & World Health Organization  & 223 & 3.1M  & Both & International & Yes \\
        WikiDoc & WikiDoc  & 33058 & 34M  & Both & International & Yes \\
        StatPearls & StatPearls  & 9.3k & 301.2k  & Both & International & Yes \\
        Textbooks & Textbooks  & 18 & 125.8k & Both & International & Yes \\
        Wikipedia & Wikipedia  & 6.5M & 29.9M  & Both & International & Yes \\
        MedCorp & MedCorp & 30.4M & 54.2M  & Both & International & Yes \\
        \bottomrule
    \end{tabular}
    \end{adjustbox}
    \label{tab:combined_sources}
\end{table*}

\section{Case Study}

\begin{longtable}{>{\centering\arraybackslash}p{6.8cm}p{6.8cm}} 
\caption{Human Annotated Sample Case.}\label{case_eurorad} \\ 
\toprule
 \multicolumn{2}{p{13.6cm}}{\textbf{Question:} A 67-year-old man with transitional cell carcinoma of the bladder comes to the physician because of a 2-day history of ringing sensation in his ear. He received this first course of neoadjuvant chemotherapy 1 week ago. Pure tone audiometry shows a sensorineural hearing loss of 45 dB. The expected beneficial effect of the drug that caused this patient's symptoms is most likely due to which of the following actions? A: Inhibition of proteasome, B: Hyperstabilization of microtubules, C:Generation of free radicals, D: Cross-linking of DNA} \\ \hline
 \multicolumn{2}{p{13.6cm}}{\textbf{Phi-4:} To determine the expected beneficial effect of the drug that caused the patient's symptoms, we need to identify the drug likely responsible for the sensorineural hearing loss and understand its mechanism of action. 1. The patient has transitional cell carcinoma of the bladder and received neoadjuvant chemotherapy. 2. A common chemotherapy drug used for bladder cancer is cisplatin. 3. Cisplatin is known to cause ototoxicity, which can manifest as sensorineural hearing loss. 4 Cisplatin will lead generation of free radicals and lead hearing loss hence the final answer should be C.} \\ \hline
  \multicolumn{2}{p{13.6cm}}{\textbf{Qwen-2.5-14B-Instruct} Forms of chemotherapy used in transitional cell carcinoma of the bladder often include platinum-based agents like cisplatin. Cisplatin is known for its ototoxicity, which can lead to sensorineural hearing loss. -> Cisplatin's mechanism of action involves the generation of free radicals, which can damage cellular components including those in the auditory system. -> Therefore, the generation of free radicals is the most likely action causing the patient's symptoms. -> The correct answer is C
}\\ \hline
  \multicolumn{2}{p{13.6cm}}{\textbf{TARSE:}} \\
  \multicolumn{2}{p{13.6cm}}{\textbf{Small Batch Finetune:}} \\
  \multicolumn{2}{p{13.6cm}}{\textbf{Pretrain Data:} MDMA: 3,4-Methylenedioxymethamphetamine (MDMA), commonly known as ecstasy (E) or molly, is a psychoactive drug primarily used for recreational purposes. The desired effects include altered sensations, increased energy, empathy, as well as pleasure. When taken by mouth, effects begin in 30 to 45 minutes and last 3 to 6 hours...} \\
  \multicolumn{2}{p{13.6cm}}{\textbf{Finetune Data:} Scenario: Dr. Smith, a seasoned urologist, is preparing for a surgical procedure to repair an anterior urethral stricture in a 45-year-old male patient. The patient is informed that a surgical resident will be involved in the operation as part of their training. The patient expresses concern about the potential impact of the resident's involvement on the surgery's...} \\
  \multicolumn{2}{p{13.6cm}}{\textbf{Inference:}
    1. Transitional cell carcinoma of the bladder -> Neoadjuvant chemotherapy -> Agent causing ototoxicity leading to sensorineural hearing loss -> Correct action of the drug is hyperstabilization of microtubules
    2. Transitional cell carcinoma of the bladder -> Neoadjuvant chemotherapy -> Ototoxicity side-effect -> Sensorineural hearing loss -> Caused by generation of free radicals
    3.Transitional cell carcinoma of the bladder -> Neoadjuvant chemotherapy -> mechanism is Cross DNA Link -> Chemotherapeutic agent causing ototoxicity -> Sensorineural hearing loss observed -> Ototoxic mechanism involves generation of free radicals -> Answer: E} \\
    \multicolumn{2}{p{13.6cm}}{\textbf{Retrieve to Verify:} 
  1. Cisplatin is a chemotherapy drug → causes ototoxicity (hearing loss) → due to accumulation of reactive oxygen species (ROS) in the cochlea → ROS cause oxidative stress and damage cochlear hair cells → Antioxidants neutralize ROS and reduce cochlear damage → therefore, antioxidants can help prevent cisplatin-induced hearing loss
  2. Chemotherapy Drug -> Cross DNA Link -> Therapy}\\ 
  \multicolumn{2}{p{13.6cm}}{
    \textbf{Final Answer:} D}\\
  \hline
  \multicolumn{2}{p{13.6cm}}{
    \textbf{Doctor Comment:}The clinical reasoning task requires distinguishing between a drug's therapeutic mechanism and its side-effect profile. While models like Phi-4 and Qwen-2.5-14B correctly identified cisplatin as the causative agent, they misattributed the drug's mechanism of action to its ototoxic side effect (free radical generation). In contrast, TARSE correctly disambiguates the therapeutic mechanism (DNA cross-linking) from side effects, demonstrating a structured reasoning pathway that mirrors human clinical logic. }\\
\\ \hline
\end{longtable}

\newpage


\end{document}